\documentclass[fleqn,usenatbib]{mnras}
\usepackage{newtxtext,newtxmath}
\usepackage[T1]{fontenc}
\usepackage{ae,aecompl}

\usepackage{graphicx}	% Including figure files
\usepackage{amsmath}	% Advanced maths commands
\usepackage{amssymb}	% Extra maths symbols
\usepackage{comment}
\usepackage{todonotes}
\usepackage{adjustbox}
\usepackage{geometry,array,float,caption}
\usepackage{subcaption}
\usepackage{etoolbox}
\usepackage{cleveref}

\definecolor{commentcolor}{RGB}{200, 10, 60}

\usepackage{scalerel}

\newcommand\reallywidehat[1]{\arraycolsep=0pt\relax%
\begin{array}{c}
\stretchto{
  \scaleto{
    \scalerel*[\widthof{\ensuremath{#1}}]{\kern-.5pt\bigwedge\kern-.5pt}
    {\rule[-\textheight/2]{1ex}{\textheight}} %WIDTH-LIMITED BIG WEDGE
  }{\textheight} % 
}{0.5ex}\\           % THIS SQUEEZES THE WEDGE TO 0.5ex HEIGHT
#1\\                 % THIS STACKS THE WEDGE ATOP THE ARGUMENT
\rule{-1ex}{0ex}
\end{array}
}
\title[Deep Learning Data Inspection for Radio Astronomy]{Deep Learning Assisted Data Inspection for Radio Astronomy}

\author[Michael Mesarcik et al.]{Michael  Mesarcik,$^{1}$\thanks{E-mail: m.mesarcik@uva.nl}
Albert-Jan Boonstra,$^{3}$
Christiaan Meijer,$^{2}$
\newauthor Walter Jansen, $^{3}$
Elena Ranguelova, $^{2}$
Rob V. van Nieuwpoort$^{1,2}$
\\
$^{1}$Informatics Institute, University of Amsterdam, Netherlands\\
$^{2}$eScience Center, Science Park 140, 1098 XG Amsterdam, The Netherlands\\
$^{3}$ASTRON, the Netherlands Institute for Radio Astronomy\\ Oude Hoogeveensedijk 4, 7991 PD Dwingeloo, The Netherlands
}

\date{Accepted XXX. Received YYY; in original form ZZZ}

\pubyear{2020}

\begin{document}
\label{firstpage}
\pagerange{\pageref{firstpage}--\pageref{lastpage}}
\maketitle
\begin{abstract}
Modern radio telescopes combine thousands of receivers, long-distance networks, large-scale compute hardware, and intricate software. Due to this complexity, failures occur relatively frequently. In this work we propose novel use of unsupervised deep learning to diagnose system health for modern radio telescopes. The model is a convolutional Variational Autoencoder (VAE) that enables the projection of the high dimensional time-frequency data to a low-dimensional prescriptive space. Using this projection, telescope operators are able to visually inspect failures thereby maintaining system health. We have trained and evaluated the performance of the VAE quantitatively in controlled experiments on simulated data from HERA. Moreover, we present a qualitative assessment of the the model trained and tested on real LOFAR data. Through the use of a na\"ive SVM classifier on the projected synthesised data, we show that there is a trade-off between the dimensionality of the projection and the number of compounded features in a given spectrogram. The VAE and SVM combination scores between 65\% and 90\% accuracy depending on the number of features in a given input. Finally, we show the prototype system-health-diagnostic web framework that integrates the evaluated model. The system is currently undergoing testing at the ASTRON  observatory. 
\end{abstract}


%%%%%%%%%%%%%%%%%%%%%%%%%%%%%%%%%%%%%%%%%%%%%%%%%%
%%%%%%%%%%%%%%%%% BODY OF PAPER %%%%%%%%%%%%%%%%%%

\section{Introduction}
Modern radio telescopes generate an ever growing amount of data. To improve spatial resolution, sensitivity, and field of view, larger telescope arrays are being constructed~\cite{Foley2016,VanHaarlem2013,Norris2010}. The increased size and capabilities of the instruments lead to more data, a higher system complexity, and ever-growing error rates.

System Health Management (SHM) is the process of detecting, diagnosing and remedying system failures to maximise system uptime. SHM and pinpointing error sources in radio telescopes today still rely on manual inspection and human interpretation of the data. This is error-prone, and large-scale spatially distributed radio telescopes such as LOFAR~\cite{VanHaarlem2013} face reliability and up-time issues. These issues stem from the scale and complexity of the systems and processing chains involved~\cite{Broekema2018}. In LOFAR for example, we expect that, due its scale and its somewhat harsh operating conditions, at any given time several components in the systems will not operate correctly. The LOFAR stations in the Netherlands are exposed to high moisture levels which may result in failure of components such as antennas and amplifiers~\cite{VanHaarlem2013}. Other reasons for failures include normal wear of components, network packets being dropped, and hardware and software errors among many others.

For the next generation exascale SKA radio telescope~\cite{Hall2007} this will be an even bigger issue. The complexity of the instruments, the myriad of observational modes, and the scale of the data transport and compute platform \cite{Jongerius2014} make accurate error detection~\cite{VanVeelen2007} and complete fault localisation very difficult. Therefore, intelligent automated SHM approaches would significantly improve the quality and availability of the observational systems. This is not only beneficial for (predictive) maintenance, operations, and cost, but it also is crucial for the science results, as accurate knowledge of the state of the telescope is essential for calibrating the system~\cite{Wijnholds2010}. 

The variability and amount of features found in data obtained from radio telescopes makes applying classic signal processing techniques for SHM difficult, as these techniques depend on specific feature morphologies \cite{Maslakovic1996,Lord1984}. Therefore, the scale, complexity and variability of features have made machine learning approaches candidate solutions to this problem. 

We contribute a deep convolutional variational autoencoder (VAE) as a solution to cope with the growing data rates and complexity of features found radio astronomical data. The proposed solution provides telescope operators with a low dimensional prescriptive space that enables the diagnosis of system health. 

Currently, research at the intersection of the fields of deep learning and radio astronomy is mostly in the categories of Radio Frequency Interference (RFI) detection \cite{Vos2019} as well as end user analysis such as galaxy morphology classification \cite{Wu2019} and detection of transient signals \cite{Connor2018,Rowlinson2019a}.

In this paper we provide an analysis of existing literature concerning the intersection of radio astronomy and deep learning. We make contributions in the form of a solution to cope with the increasing data volumes while making it more feasible to still have expert knowledge in the loop. The system presented in this work can assist with data inspection, and is capable of detecting anomalies that can be used in diagnosing system health. It is being used in production for the LOFAR telescope. 

Furthermore, a discussion and analysis of the preprocessing techniques that are applied to the spectrograms obtained from the LOFAR are discussed as we found this far from trivial, and has significant impact on the quality of the end results. Finally, we establish performance evaluation metrics that allow quantitative and qualitative comparisons of different approaches and implementations.

This paper begins with the analysis of existing literature concerning the intersection of radio astronomy and deep learning in Section~\ref{sec:Related_Work}. In Section~\ref{sec:processing} we discuss and analyse the preprocessing techniques that we apply to the spectrograms obtained from LOFAR and the data obtained from the HERA simulator. Section~\ref{sec:Architecture} documents the deep learning architecture used, while Section~\ref{sec:evaluation} establishes performance evaluation metrics. Finally, in Section~\ref{sec:Results} we present results and in Section~\ref{sec:Conclusions} the conclusions are given.

\section{Related Work}
\label{sec:Related_Work}

The adoption and penetration of deep learning techniques in radio astronomy have been limited, but efforts are rapidly increasing. Progress has been made in specific areas such as galaxy morphology classification, RFI, transient and context-specific anomaly detection. However, to the best of our knowledge there have been no attempts to use deep-unsupervised learning techniques for the evaluation of data quality in radio telescopes. In this section, we sketch the landscape of research relating to the intersection of machine learning and radio astronomy.

\subsection{Radio frequency interference mitigation}
\cite{Akeret2017} describe and test an implementation of a U-net architecture~\cite{Ronneberger2015} for purposes of RFI mitigation. The system is trained on the magnitude component of spectrograms obtained from the HIDE \& SEEK simulator~\cite{Akeret2017b} and tested on both simulated and real data obtained from the Bleien Observatory. 

In similar work~\cite{Kerrigan2019} use a modified U-net architecture that classifies RFI on a pixel-per-pixel basis for a given spectrogram. Notably, the paper shows that the use of both magnitude and phase information yield improved performance in comparison to a magnitude-only network for both the use of simulated and real data obtained from the HERA telescope~\cite{DeBoer2017}. This work is further extended in work by \cite{Yang2020} through their addition of residual connections between layers which increases prediction accuracy as well as the speed of training convergence.  

Our research generalises work by~\cite{Kerrigan2019} through considering RFI as a subset of the features found in the LOFAR data. Using this premise, we document an extension of the performance analysis when phase information is included in this work. 

\subsection{Fast radio bust and transient detection}

In work by \cite{Connor2018} it is shown that deep convolutional neural networks are effective in the detection of candidate Fast Radio Bursts (FRBs). As the number of training examples is very low, the authors superimpose various FRB models onto spectrograms obtained from CHIME path finder \cite{Amiri2019} and from the Apertif telescopes \cite{Oosterloo2009}. 

This work leverages an ensemble of models which are trained separately on multiple observation domains, these being the time frequency pulse profiles, the collapsed pulse profiles, the dispersion measures in time and the sky beam configurations.  

In work by \cite{Agarwal2019} this principle is extended to out-the-box solutions for existing network architecture where transfer learning is used to combat the number of required FRB training instances. In contrast to \cite{Connor2018}, this work does not make use of time series data and it is shown that their models obtain slightly higher accuracies. 

In more general transient detection cases \cite{Rowlinson2019a} show that through the combination of statistical thresholding mechanisms and traditional machine learning techniques, that transient detection is possible. This work utilises a high degree of domain knowledge and is not particularly flexible in other anomaly detection contexts. 

Corollary, in our work we propose a simple preprocessing pipeline in conjunction with a model requiring very little domain knowledge to understand and operate. Furthermore, as we heavily down-sample the time-frequency spectrograms from LOFAR it is most likely that FRB detection is not possible. Finally, we do not investigate transfer learning; we will experiment further with this in future work.

\subsection{Radio frequency anomaly detection}
In work by~\cite{OShea2016} various deep learning architectures are shown to yield increased performance over traditional methods such as Kalman novelty detectors~\cite{Vittaldev2012} in the context of radio frequency anomaly detection. This work uses naive implementations of fully connected, convolutional and Long Short-Term Memory (LSTM) based networks to detect known, superimposed anomalies in spectrograms for various frequency bands. An unsupervised LSTM-based extension of this work is described by~\cite{Tandiya2018}, where Prednet~\cite{Lotter2016} is used for frame sequence prediction. It is applied to predict the next frame in a sequence of processed time-frequency spectrograms and is shown to yield increased performance to the work done by~\cite{OShea2016}.

In work applied to radio astronomy and the Search for Extraterrestrial Intelligence (SETI), \cite{Zhang2019} show that an LSTM-based sequence predictor is suitable in the SETI context. This is achieved from the temporal methodology used to distinguish terrestrial RFI to extraterrestrial techno-signatures. This technique is leveraged by using a generative LSTM to predict what the output should be when pointing the telescope at a particular source. If the prediction is highly dissimilar to the actual observation then an anomaly is flagged. 

Although these works show the relevance of detecting anomalies in sequence based scenarios, they are unable to offer low dimensional projections of the input data. For this reason our work focuses on convolutional autoencoders for low dimensional visualisations. In future work we aim to investigate the possible integration of sequence based models for anomaly detection. 

\subsection{Galaxy Classification}
There have been multiple attempts, with varying success, to improve performance and accuracy of supervised neural networks for the tasks of supervised galaxy classification~\cite{Mingo2019,Lukic2018,Wu2019,Aniyan2017}. In work by \cite{Ralph2019} it is shown that unsupervised learning techniques are effective in data exploration and the clustering of the morphologies of radio galaxy images. This is achieved through the use of an autoencoder (AE) to reduce the dimensionality of the input data and ensure that their compressed representations are affine-transform invariant. The compressed representations are then used to train a Self-Organising Map (SOM) to create a similarity matrix of the input data. Finally, k-means clustering is used to assign cluster labels with the associated learnt similarity maps. However, this work lacks evaluation of the degree to which their AE ensures that the projection is affine invariant. This lack of evaluation propagates into the results of their work, where different rotations galaxy morphologies are clustered separately. 

In contrast, in our work, we show that a Variational Autoencoder (VAE) is capable of creating meaningful low dimensional projections of astronomical data without the addition of a SOM. In later sections, we describe the affine invariance properties of the VAE used. We show that increasing the latent vector's dimensions results in the increased ability of the model to generalise to affine transformed augmentations of a particular input feature. 

\section{Data Preparation and Preprocessing}
\label{sec:processing}
\begin{figure*}
    \centering
    \includegraphics[width=\textwidth]{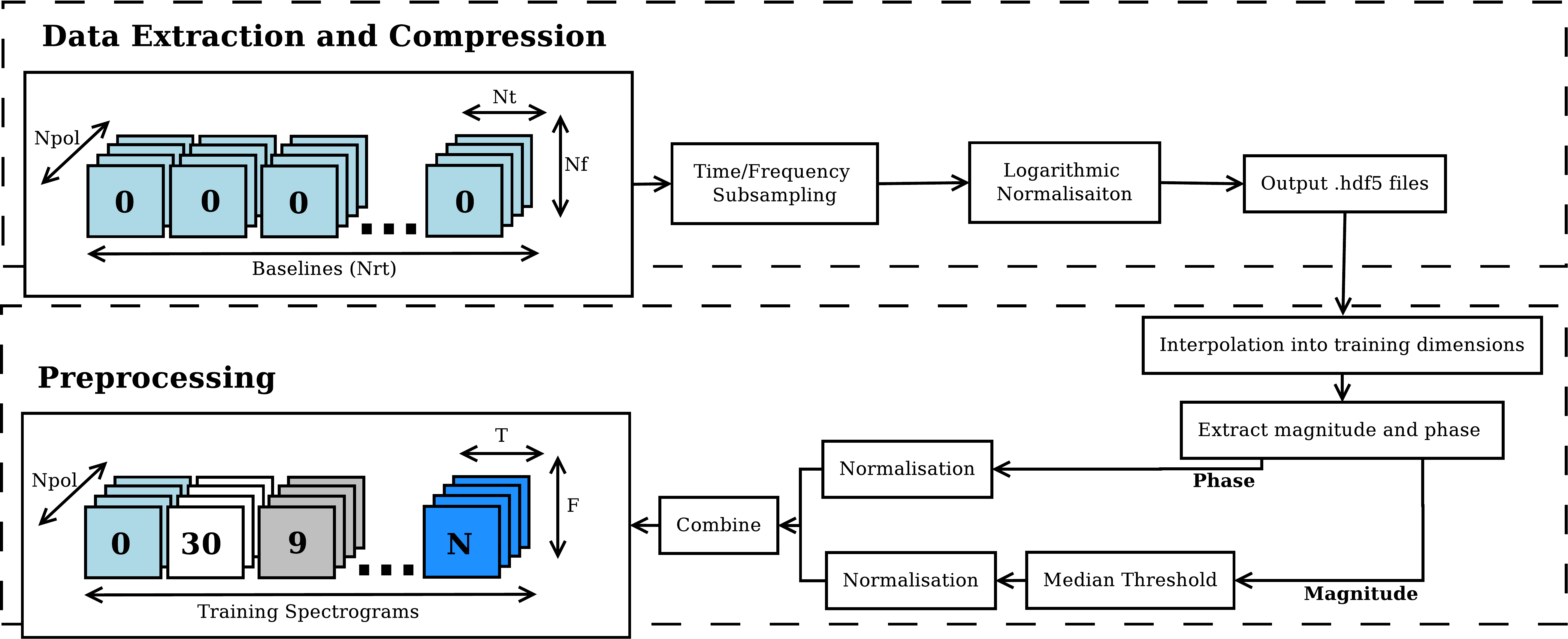}
    \caption{Block diagram showing the data extraction and preprocessing pipelines applied to the observations before training}
    \label{fig:preprocessing}
\end{figure*}

In this section, we discuss and analyse the preprocessing techniques that we apply to the spectrograms obtained from LOFAR. This step is often under-emphasised or even omitted in papers describing the application of machine learning to radio astronomy based applications. This is unfortunate, since we found this is far from trivial, and has significant impact on the quality of the end results and is important for the reproducibility of the research.

Preprocessing is particularly challenging for radio astronomy compared to other fields. Examples of some of the preprocessing challenges faced in the field are the huge 5-dimensional data cubes, that may lead to issues with down-sampling due to the high data rates; the complications of having amplitude and phase information, which must be treated separately; and the extremely noisy data with a very high dynamic range (due to interference).

\subsection{Data extraction, down-sampling and reformatting}

The LOFAR radio telescope \cite{VanHaarlem2013} creates very wide-field sky images using aperture synthesis, and provides coherently added time series signals for transients research. The telescope consists of $N_{rt} = 51$ phased array stations distributed over eight European countries, with its core located in the Netherlands. Each station consists of an array of 96 dual polarisation LBA antennas (10-90~MHz) and 48 or 96 dual polarisation HBA antenna tiles (110-250~MHz). Antenna signals are added coherently in each station beamformer for each sub-band of approximately 200~kHz, allowing multiple parallel beams to be created. Beamformed station signals are transported to the central processor and mutually correlated to form a correlation matrix with a minimum channel width of about 0.7~kHz. This set of correlations are called visbilities. After calibration, the visibilities are converted to sky maps by using 2D Fourier transforms and related techniques.

Before applying final calibration steps and providing the data to the astronomers and the archive, the visibility data and meta-data are inspected and annotated. The visibility data hyper cube is four dimensional (or five dimensional, depending on perspective) with each dimension corresponding to time, frequency, polarisation, and baseline. The correlation integration time and observation length determine the number of time 'snapshot' samples denoted by $N_t$; the number of frequency bins is denoted by $N_f$. Given a dual polarisation system, the number of polarised correlation products $N_{pol}$ usually is 1, 2 or 4. Each visibility is a correlation product between two telescope station signals. The separations between pairs of stations are called baselines and are represented by vectors. With $N_{rt}$ telescope stations, there exist $N_{rt}$ auto-correlations and $\frac{1}{2} N_{rt}(N_{rt}-1)$ cross correlations for each polarisation combination ('xx', 'xy', 'yx', and 'yy'). If the baselines are considered to represent one dimension, then LOFAR data cubes are four-dimensional, if the baselines are considered to lie in a 2-D aperture plane, then the hyper-cube can be considered five-dimensional.

A LOFAR data set can be relatively large. The data size of an aperture synthesis observation is given by $\frac{1}{2} N_t N_f N_{rt}(N_{rt}+1) N_{pol}N_{bits}$. Given a $\Delta T = $ 10 hour observation duration with $\tau_{int}$ = 1 second integration time, 50 MHz bandwidth and 1 kHz channel resolution this leads to a data set size in the order of 100~TBs. Each visibility contains $N_t \times N_f$ complex data points. For the parameter values mentioned above this means about $2\times10^9$ complex data points for each of the 1275 baselines for each polarisation. 

As this is rather high, data inspection is usually performed using a lower spectral resolution, and/or time averaging, and/or sub-sampling. The so-called \textit{inspection plots} are based on either low spectral resolution auto-correlation spectra created at the stations, or on compressed visibility data. For each LOFAR aperture synthesis observation, an additional compressed \texttt{.hdf5} dataset is produced in parallel to the full measurement set. The compression parameters are tunable, but a 100~TB observation is typically reduced to order 1~GB. These compressed data sets are input to the deep learning networks.

The data compression stages can be seen in the top half of Figure~\ref{fig:preprocessing}. The \texttt{.hdf5} file data compression includes the following steps. First, the the dataset is sub-sampled in time, using a regular grid. The next step is aimed at reducing the dynamic range of the data so that it fits in fewer bits. This includes taking the logarithm of the absolute value of visibilities, and normalisation by a scaling factor computed as the maximum value for each baseline, sub-band, and polarisation. The normalisation factor is stored in the \texttt{.hdf5} file so that the original data can be reconstructed albeit with reduced resolution. The compressed data are stored as 8 bit integers for the real and imaginary parts of the complex visibilities.

As each observation contains thousands of these visibility spectrograms, deep learning approaches may alleviate the burden of manual data quality inspection.

\subsection{Training set preprocessing}
The main considerations for the preprocessing of the time-frequency spectrograms that are used for training are the conversion to the appropriate domain, the normalisation method, the correct scaling of features and size of each spectrogram. The preprocessing scheme used in this work can be seen in the bottom section of Figure~\ref{fig:preprocessing}. 

As shown in work by~\cite{Kerrigan2019}, the use of both amplitude and phase components of the complex visibilities yield an increase in performance in the classification of RFI. For this reason, we deem it critical to make use of both components of the complex visibilities. The model evaluation shown in Section~\ref{sec:evaluation} demonstrates the differences in performance when changing the domain of the training data. 

Due to the memory and storage constraints whilst training, all complex spectrograms are constrained in size. As the extracted \texttt{.hdf5} could consist of different observation durations each with a different number of sub-bands, it was necessary to resize the visibility matrices. The resizing was performed by down-sampling all dynamic spectra greater than 128x32 and interpolating all spectrograms smaller than 128x32, in frequency and time. 

% Why did we choose this particular size? can you say something about using other sizes? You did try smaller spectograms in the beginning, right? 

To increase the dynamic range of the astronomical features present in the magnitude component of the complex visibilities, we perform naive RFI suppression. We achieve this through the use of a median based thresholding mechanism which we only apply to the magnitude component of the spectrograms. We don't apply the scheme to the phase components of the visibilities, as they are bounded between $-\pi \leq \phi \leq \pi$. 

The median-threshold was calculated on a per-spectrogram basis so that every different baseline had a unique threshold such that 
\begin{equation}
    |V_{i,j}'(\tau,\nu)| = 
     \begin{cases}
          \sigma, \text{if}\quad |V_{i,j}(\tau,\nu)| \ge \sigma \\
          |V_{i,j}(\tau,\nu)|, \text{otherwise} 
     \end{cases}
\end{equation}
where $\sigma$ is the first standard deviation of the visibility of antennas $i$ and $j$ given by $ V_{i,j}(\tau,\nu)$ where $\tau$ and $\nu$ correspond to the time and frequency dimensions respectively.

As a consequence of the magnitude being unbounded, we normalised the magnitude between 0 and 1 on a per baseline basis after the naive RFI suppression. Additionally, to ensure that the phase component was weighted equally in training, we normalise the phase component to the same scale as the magnitude. 

\begin{figure}
    \centering
    \includegraphics[trim={0 2.2cm 0 2.2cm},clip,width=0.9\linewidth]{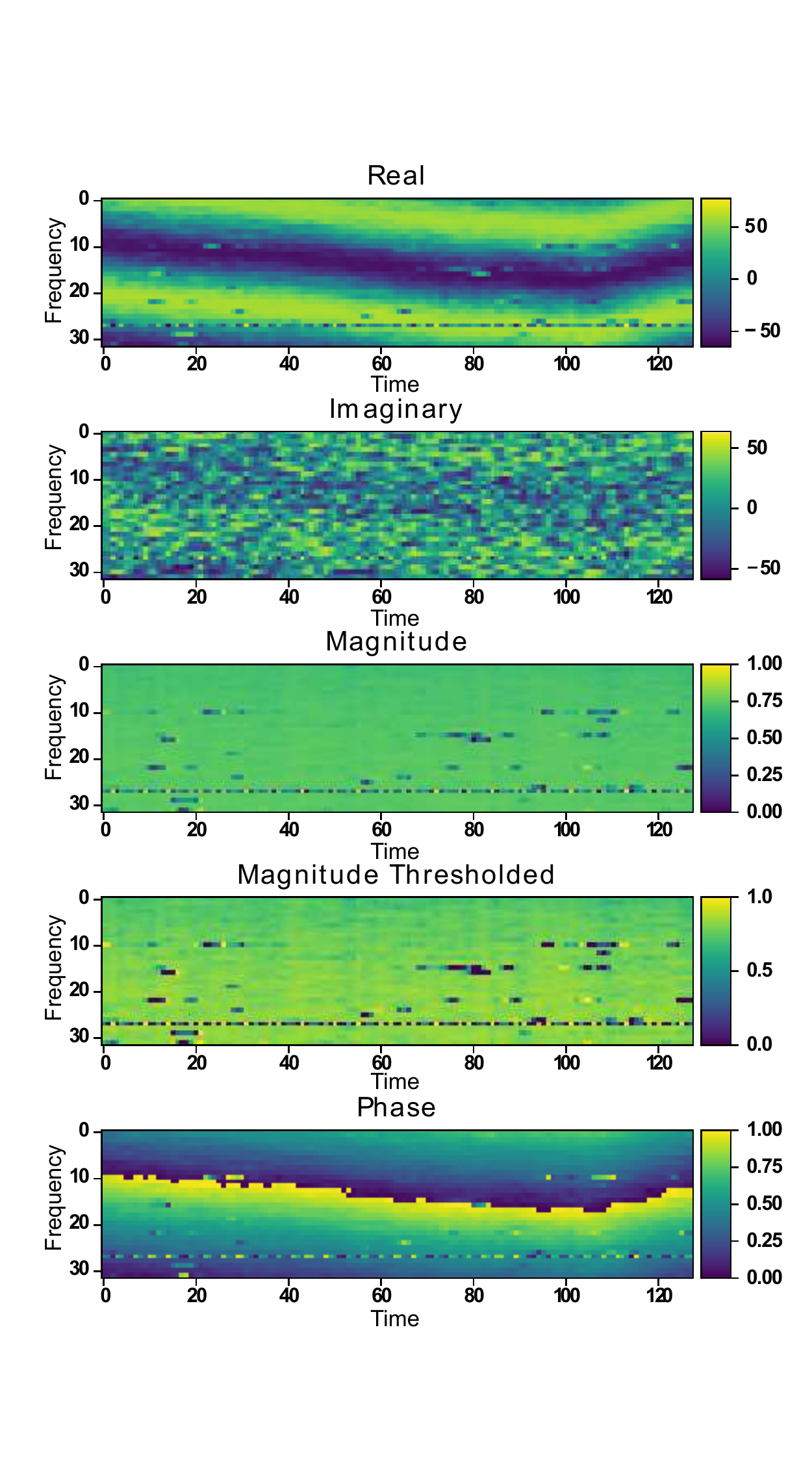}
    \caption{The preprocessing stages applied to a randomly selected baseline from the LOFAR training set.}
    \label{fig:processed_baselines}
 \end{figure}

An illustration of the stages of the preprocessing pipeline can be seen in Figure~\ref{fig:processed_baselines}.

\section{Deep Learning Architecture}
\label{sec:Architecture}
\begin{figure*}
    \centering
    \includegraphics[width=\textwidth]{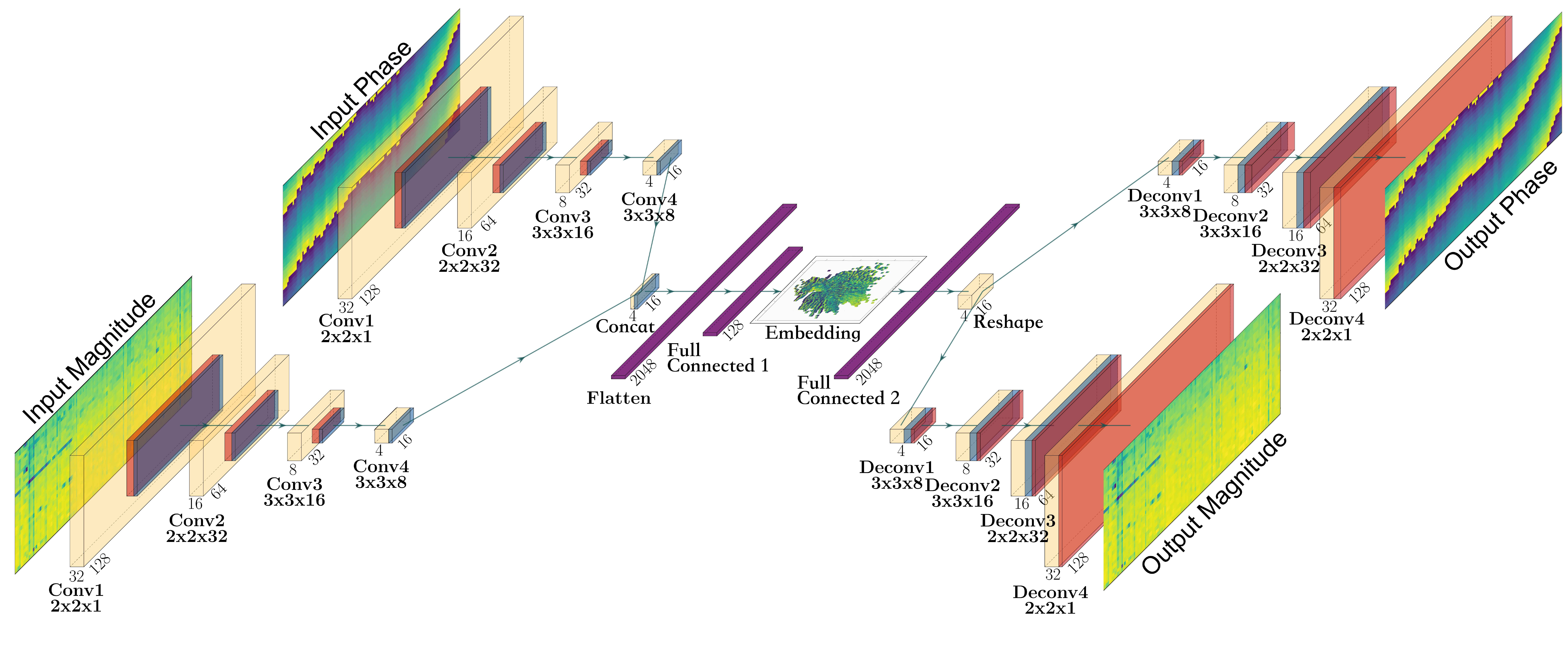}
    \caption{The architecture of the magnitude and phase-based VAE used for the deep embedding of the astronomical spectrograms. The yellow layers correspond to convolutional layers, the green and blue layers correspond to max pooling and upsampling layers respectively, whereas the mauve layers correspond to batch-normalisation layers. For the intermediate representation, purple is used to designate the flattening operations, fully-connected layers as well as the latent representation.}
    \label{fig:architecture}
\end{figure*}

In this section we describe the design parameters and constraints of the implemented deep learning architecture. It was found that the inclusion of both the magnitude and phase components into the model required consideration to be given to normalisation conditions due to the dynamic range of some features in magnitude.  Figure~\ref{fig:architecture} shows the magnitude and phase VAE-based architecture used in this work.

\subsection{Architectural overview}
An autoencoder is a neural network architecture used to learn compressed representations of data in an unsupervised manner. The network can be considered in two parts, the encoding network, $h = f(X)$, that generates an intermediate representation of the input data, $h$, and a decoding network, $r = g(h)$, that regenerates the input data from the encoded representation. Commonly, the interconnection between the encoder and the decoder networks is referred to as an embedding as it has a lower dimensionality than input, thereby enabling a compressed representation of the data~\cite{Goodfellow2015}.

% If we keep this level of detail, we need to explain and/or cite a bit more here I fear: L1 and L2 loss functions KL-divergence
VAEs combine variational inference techniques~\cite{Blei2017} with neural networks that are used as complex transform approximators. Unlike traditional autoencoders that use $\mathcal{L}_1$ or $\mathcal{L}_2$ loss functions, variational autoencoders try to minimise the KL-divergence between the input distribution, $q(X)$, and the output distribution, $p(X)$, given the intermediate representation, $z$, as given by

\begin{equation}
    \mathbf{min}\; \mathcal{D}_{KL}(q(z|X)|| p(z|x)) = \sum_z q(z|x) \log \Big( \dfrac{q(z|X)}{p(z|X)} \Big).
    \label{eq:KL}
\end{equation}

The rearrangement of the minimisation problem in \ref{eq:KL}, shows that rather than minimising the KL-divergence it is possible to maximise the Evidence Lower Bound (ELBO) \cite{Goodfellow2015} as a good approximation to the solution. This is achieved through the use of an optimisation such as stochastic gradient descent (SGD) with back propagation in order to maximise the lower bound of $\log{P(X)}$, such that the adjusted function to maximise is given by 

\begin{equation}
    \mathbf{max} \; \log(P(X)) \geq \mathbb{E}_{q(z|X)} \big[ log(p(X|z)\big] - \mathcal{D}_{KL}(q(z|X) || p(z)).
\end{equation}

When we parameterise $p(z)$ by $\theta$ and $q(z|X)$ by $\phi$ the loss function of the VAE can be explicitly stated as

\begin{equation}
    \begin{split}
        \mathcal{L}_{\mathrm{ELBO}}(\theta, \phi, X) = - \mathcal{D}_{KL}(q_\phi(z|X) ||& p_\theta(z)) \\
        &+ \mathbb{E}_{q_\phi (z|X)} \big[ log(p_\theta(X|z)\big]  .
    \end{split}
\end{equation}

In the case of the autoencoder-based structure, the $q_\phi(z|X)$ term may be considered as the encoder, mapping the input distribution $q_\phi(X)$ to the latent projection given $p_\theta(z)$, whereas the $p_\theta(X|z)$ term may be considered the decoder, mapping the output distribution $p_\theta(X)$ given $z$~\cite{Min2018}.

Typically, the prior distribution, $p_\theta(z)$, is obtained by sampling it from a Gaussian. However, as the sampling operation cannot be used to back-propagate a gradient (as it is not differentiable) a \textit{reparameterisation trick} is used. The reparameterisation removes the non-differentiable sampling operations from the network and replaces them with differentiable operations. More details regarding the variational autoencoder may be found in~\cite{Kingma2013a}. 

In the VAE architecture used in this research, convolutional layers are used for both the encoding and decoding networks. Convolutional neural networks (CNNs) are a class of Artificial Neural Neworks (ANN) used to preserve spatial dependence in image processing tasks. They differ from classical, fully-connected neural networks by the use of spatially independent trainable filters, instead of fully connected neurons, and are used for the detection of spatial features. This makes them applicable in our context as astronomical and terrestrial sources are spatially localised to particular regions in the radio astronomical data. The convolutional layers shown in Figure~\ref{fig:architecture} are represented in yellow. 

Structurally, CNNs consist of convolutional layers followed by pooling layers. Commonly, max-pooling layers are used in CNNs. These layers compute the maximum value in a sliding-window applied to the output of each convolutional layer. This decreases the size of the output from each layer as well as selecting the salient features apparent in each sequentially decreasing convolutions layer. The outputs of the pooling layers are fed to non-linear activation functions which produce spatial activation maps that describe the per-pixel response of each convolutional layer to their inputs~\cite{Lecun1998}.  

A structural consequence of the autoencoder-based architecture is that consecutive layers in the encoder network should decrease in size. This means that, for the decoder to regenerate the input data, upsampling must be applied to the encoded representation. The effect of this is that after each convolutional layer of the decoder a upsampling layer is added to restore the correct dimensionality of the output. In the case of the architecture used, the number of filters was decreased by a factor of 2 for each sequential convolutional layer in the encoder, and is up-sampled by a factor of 2 in each consecutive convolutional layer in the decoder. 
In Figure~\ref{fig:architecture} the pooling layers are shown in green, whereas the upsampling layers are shown in blue. 
% I don't really see these colors?

\subsection{Architectural motivation and constraints}
The major consideration that we made regarding the VAE architecture was the structural segmentation between the magnitude and phase components of the complex visibilities. We deem it critical to make use of both components as they contribute differing representations of particular astronomical and terrestrial phenomena. In section~\ref{subsec:accuracy}, we discuss a performance comparison of network architectures that use different combinations of real, imaginary, magnitude and phase components.

\begin{table}
	\centering
	\caption{The parameters of VAE that were determined through a coarse grid search over the hyperparameter space.}
	\label{tab:training_paramters}
	\begin{adjustbox}{width=\columnwidth}
    	\begin{tabular}{lccr} % four columns, alignment for each
    		\hline
    		\textbf{Parameter type} & \textbf{Value}\\
    		\hline
    		Optimiser & ADAM\\
    		Learning rate & 0.0001 \\
    	    Loss function & Binary Cross Entropy \\
    		Batch size& 256\\
    		Number of epochs & 200 \\
    		Activation functions & Relu\\
    		Batch normalisation momentum & 0.99 \\
    		\hline
    	\end{tabular}
	\end{adjustbox}
\end{table}

A result of using the magnitude and phase components for training the VAE-based model is that the learnt representations from both the phase and magnitude encoders need to be joined together. As features appear differently in each domain, we determined that normalisation after each convolutional layer was necessary. The normalisation ensured that the independent magnitude and phase activations are maintained between 0 and 1, so that no higher magnitude activations, such as RFI, may take precedence over the learnt representations. Commonly, the normalisation is performed in batches. They can be seen in mauve in Figure~\ref{fig:architecture}. 
Similarly to~\citeauthor{Kerrigan2019}, we found that the concatenation of the learnt normalised magnitude and phase activations yields increased performance. In the case of the VAE architecture, we concatenate the magnitude and phase components after the last convolutional layer of the encoder such that their joint-embedding could be determined. 

As the objective of this work is a data inspection and visualisation tool, it was necessary to reduce the dimensions of the input spectrograms to 2 dimensions. This is critical, as visualisation and navigation of a two dimensional space is intuitive for the end users of this system at observatories. The impact of this is that the 2-dimensional embedding of the 128x32 sized spectrograms results in a 97\% reduction in dimensionality which limits the VAE's reconstruction abilities. In other literature that use autoencoders for dimensionality reduction~\citeauthor{Guo,Ralph2019} and representation learning~\citeauthor{Wu2019},higher dimensional latent spaces are often used to encode high-order features such as affine transformations. However, in our work, the latent representation is limited to 2-dimensions.  The effect of this is that some higher-order features are lost and cannot be regenerated by the decoder network. 

A benefit of the architecture we use is the ability of the VAE to generate new samples from the learnt distribution of the complex radio astronomical spectrograms. The ability of VAEs to generate new labelled data is described in more detail in~\cite{Pu2016}. Although the generative aspect of these networks are not used in this work, it is an interesting consideration as this model can be used to generate new labelled training data given that the latent projection can be labelled. We intend to investigate if we can exploit this feature to automatically classify anomalies in future work as described in \cite{Akcay}.

To ensure the selection of the optimal hyper-parameters of the model, we performed a coarse grid search  over the optimiser, loss function, filter sizes, batch size, activation functions and layer configurations. The selected hyper-parameters are shown in Table~\ref{tab:training_paramters}.

\section{Model Evaluation}
\label{sec:evaluation}
A critical aspect of the development of unsupervised embedding algorithms is the ability to validate the model's performance. Unsupervised model evaluation is an implicitly challenging task as the training data is unlabelled. Therefore, it is hard to verify whether the inputted data is being correctly represented. 

In our research, challenges arise from the compounding of combinations of features present in a spectrogram. For example, a given spectrogram from LOFAR may contain several features such as narrow band RFI, point sources, solar storms, Cassiopeia in the antenna side-lobes and many more in a particular spectrogram. This means that for a single observation the total number of combinations of features is given by  
\begin{equation}
    n_c = \sum_{k=1}^n \dfrac{n!}{(n-k!)k!} +1
    \label{eq:explosion}
\end{equation}
where $n$ is the total number of possible features, and $n_c$ designates the number of different clusters, each with a particular combination of astronomical features. In the case of 10 features, the number of combinations of features will yield 1024 different classes, given that the $\emptyset$-set is included. 

To create a controlled environment to evaluate the embedding, we use a radio astronomy simulator so that the number of features present in a particular spectrogram could be designed in a predictable manner. Additionally, the simulator allows feature labels to be generated with each spectrogram, such that the accuracy of the separation of the features can be measured, allowing a quantitative analysis. We later perform a manual qualitative analysis with real LOFAR data.

\subsection{Quantitative evaluation using simulated data}
We selected the HERA~\cite{DeBoer2017} radio astronomical data simulator, as it is capable of generating various astronomical and terrestrial events. Features such as point sources, pseudo-sky models and various forms of RFI can be generated. 

Through the use of the simulator, we generated 20,000 training spectrograms with multiple baselines of a \textit{pseudo-sky model}. We added random combinations of different features to this. These features were: point sources, narrow and broadband RFI, gain fluctuations and antenna cross talk. The features and their parameters that we used can be seen in Table~\ref{tab:simulator_parameters} and the exact model specifications of these features are described in work by~\cite{Kerrigan2019}. 

The data obtained from the HERA simulator has far fewer features and significantly smaller variability of parameters of those features compared to the real-world LOFAR data. Therefore, the results from the simulated data do not guarantee performance in geometric-separability when applying the same model to the LOFAR data. However, by evaluating the model performance both prescriptively on the HERA data and descriptively on the LOFAR data, it is possible to ensure a degree of confidence regarding the obtained results. In addition to this, the use of both synthesised HERA data and real-world LOFAR data prove the generalisability of the model for different astronomical instruments and data sets. 

We used the simulated data to train the VAE. We generated an embedding that we evaluate using a linear Support Vector Machine (SVM) classifier~\cite{Bracke2001}. Linear SVMs are a \emph{supervised} machine learning technique that classifies labelled data by segmenting the N-dimensional mappings using hyper-planes.  The classification output generated from the SVM is used as an evaluation metric of the VAE's embedding performance, because the classification accuracy increases proportionally to the euclidean separation of spectrograms with the same multi-feature labels in two dimensional space~\cite{Mavroforakis2006a}.

Since the focus of this work is on visualisation of high dimensional spectrograms, and not on classification, we chose a na\"ive classifier. Naturally, the choice of a more sophisticated classification technique would yield improved performance in many cases. However, by illustrating the effectiveness of the na\"ive classifier, we prove that the VAE is clearly capable of separation of features in the low dimensional embedding. 

\begin{table}
	\centering
	\caption{Parameters of the HERA simulator used to synthesise training data.}
	\label{tab:simulator_parameters}
	\resizebox{\columnwidth}{!}{
	
    	\begin{tabular}{lccr} % four columns, alignment for each
    		\hline
    		 \textbf{Feature Type}& \textbf{Value}\\
    		\hline
    		    Baseline geometric delays & 5~ns, 10~ns, 20~ns, 100~ns, 500~ns \\
    		    Number of frequency channels & 64 \\
    		    Number of time samples & 128 \\
    		    Number of uncorrelated sources & 200 \\
    		    Synthesised noise type & Gaussian \\
    		\hline
    	\end{tabular}
    	}
\end{table}

\subsection{Qualitative evaluation using LOFAR data}
To demonstrate the real-world efficacy of the model, we evaluate the low dimensional embedding descriptively using data obtained from the LOFAR telescope. A prescriptive evaluation was not possible on the LOFAR data embedding as there exist a multitude of features that can be simultaneously present in a given spectrogram. Moreover, these features can vary significantly across different baselines. The variability of features is a result of the geometric localisation of the terrestrial or astrophysical phenomena to a particular station. The effect of this is that localised sources will appear significantly different in geographically distant stations. Consequentially, different labels would have to be given to the same feature for different baselines which presents a problem when trying to evaluate separability of features in a prescriptive evaluation of results. Additionally, the effort of manually labelling enough data with many combinations of features is prohibitive.

For this reason and the compounding of features shown in Equation~\ref{eq:explosion}, traditional clustering evaluation methods such as accuracy (ACC) and normalised mutual information (NMI) \cite{Min2018} are not suitable metrics in this context. We aim to perform further analysis regarding anomaly and outlier detection in future work.

We evaluate the LOFAR results through visual inspection of the embedding plots, scatter plots with the magnitude spectrograms superimposed on the points, as well as considering the difference in the generative abilities of the input and output spectrograms. 

To evaluate the LOFAR-based results, we trained the VAE-based model using 327 unique observations, each consisting of between 600 and 3000 baselines from both the LBA and HBA stations. From these observations, 256 baselines were randomly sampled from each file. We preprocessed the data using the preprocessing pipeline described in Section~\ref{sec:processing}. The result of this was that we trained the model using 60672 32x128-sized complex spectrograms.

Finally, we integrated the embedding plots into a data inspection environment used by the data commissioning astronomers at the ASTRON observatory \footnote{https://www.astron.nl/}. With the use of experts in the loop, the system was iteratively evaluated whilst in the development phases.
%%%%{\color{red} I dont think this really adds anything so i removed it, do you all agree?  The interface of the data inspection tool used at the observatory can be seen in Figure~\ref{fig:LOFAR_results}}.

% general remark: We have two options for the structure of the results. What we have now: 
% eval Hera + eval LOFAR
% results Hera + results LOFAR
% the alternative would be:
% eval Hera + results Hera
% eval LOFAR + results LOFAR.
% I have a weak preference for the second option I think, this nicely separates the simulated results, and the work on the real data. Then, the description of the actual tool (interface) follows naturally...
% let's discuss tomorrow...

\section{Results}
\label{sec:Results}
In this section, the effectiveness of the designed variational autoencoder is shown. This is done by evaluating the model's ability to give meaningful low dimensional projections for telescope operators diagnosing system health. In all cases the training data consisted of 80\% of the available data and the remaining 20\% was used for testing. For the quantitative evaluation of the HERA data, the model's classification accuracy is used as the primary performance metric unless otherwise specified. A portion of the training set from the LOFAR telescope may be found in~\cite{Mesarcik2020a}.

\subsection{Simulated data}
\subsubsection{Model Accuracy}
\label{subsec:accuracy}
As a first stage approach to determine in which domain the model performs best, we tested six different network architectures. These being a VAE trained on the real component only, the imaginary component only, both the real and imaginary components, on the magnitude only, the phase only and on both the magnitude and phase components of the complex data. From this, we evaluated the classification performance for each of the different architectures.

\begin{figure}
    \centering
    \includegraphics[width=\columnwidth]{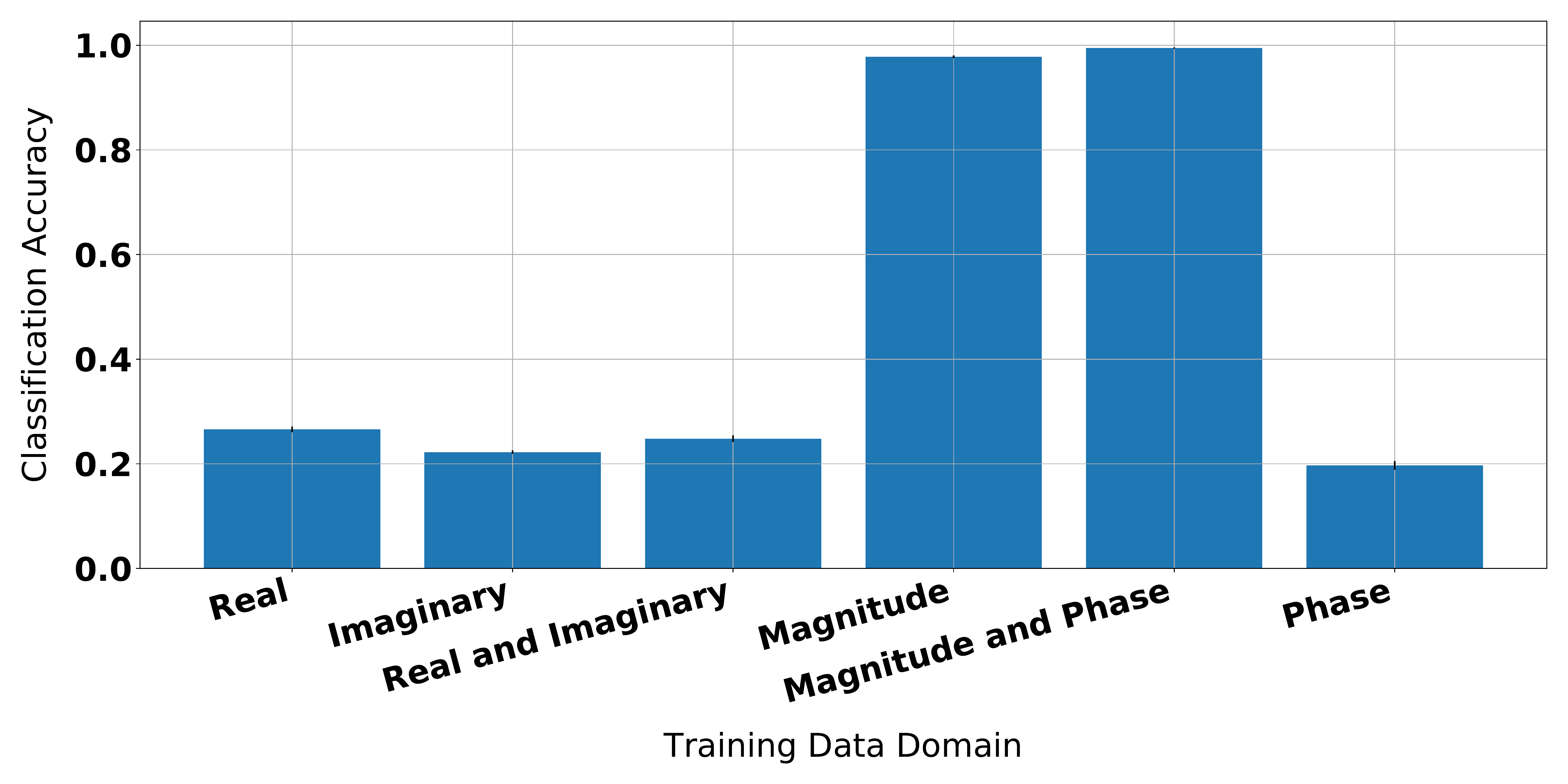}
    \caption{Classification accuracy of the na\"ive SVM classifier applied to the embeddings when the VAE was supplied data with 5 different features presented in different domains.}
    \label{fig:SVM}
\end{figure}

It is clear from the results shown in Figure~\ref{fig:SVM} that using both the magnitude and phase components of the complex visibilities yields the best classification accuracy. Albeit a small performance increase, in comparison to using the magnitude component only, it is comparable to those results reported by \cite{Kerrigan2019}. Furthermore, even though the real and imaginary representation contains the same information as the magnitude and phase representation, using the latter shows a significant performance gain. This can be explained by the complex time-frequency data being represented in a more interpretable manner for the convolutional layers of the VAE. For this reason, the magnitude and phase-based network architecture is used from this point onward in the experiments of this paper. 

\begin{figure}
    \centering
    \includegraphics[width=\columnwidth]{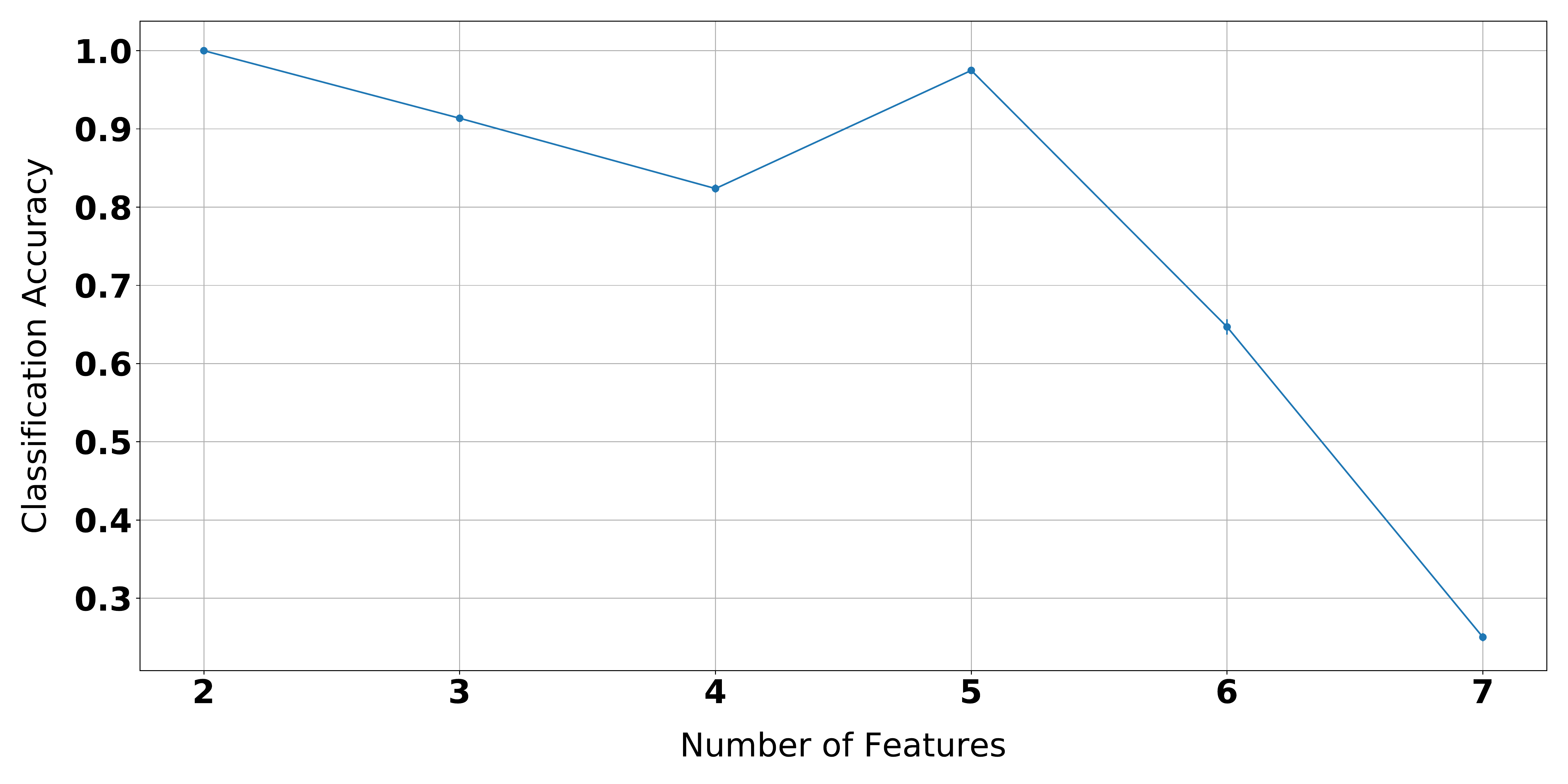}
    \caption{Classification accuracy of the na\"ive SVM classifier applied to the embeddings while varying the number of features in the HERA simulator. }
    \label{fig:Features_tests}
\end{figure}

To draw performance comparisons between the models trained on HERA and LOFAR data, we ran experiments to determine the HERA-trained model's classification accuracy as the number of features increases. We did this by varying the number of features, $n$, in a given HERA data training set from 3 to 7 and calculating the classification accuracy of the model. In Figure~\ref{fig:Features_tests} we can see that as the number features increases, the classification accuracy of the network decreases significantly.    

There is a significant classification performance loss from 5 features onward, however the use of only classification accuracy as an evaluation method does not perfectly capture the performance of the model. The linear SVM-based classifier's performance requires the embedded spectrograms to be linearly separable in their latent projection, however it shows that as the number of features increases the linear separability decreases. This is due to the fact that pronounced features such as persistent narrow band RFI tend to take precedence over lower power features such as astronomical sources such that when compounded, features consisting of a weak and strong feature are localised to the same embedding location.

This being said, when visually inspecting the embeddings obtained from models trained on data with a high number of features it is clear that certain features are well separated. This result is illustrated in \Cref{fig:Embedding_overlay,fig:Embedding_colours} where the top-rightmost cluster corresponds to all spectrograms that contain narrow-band radio station based RFI but no radio astronomical sources. Conversely, the top leftmost cluster contains narrow-band radio station based RFI and source structures. This is an important result that shows that even though the SVM classification accuracy is poor, the model correctly embeds the compounded features in a hierarchical manner that are easily interpreted by human operators.

\subsubsection{Feature Separability}
\label{subsec:feature_sep}
\begin{figure}
    \centering
    \includegraphics[width=\columnwidth]{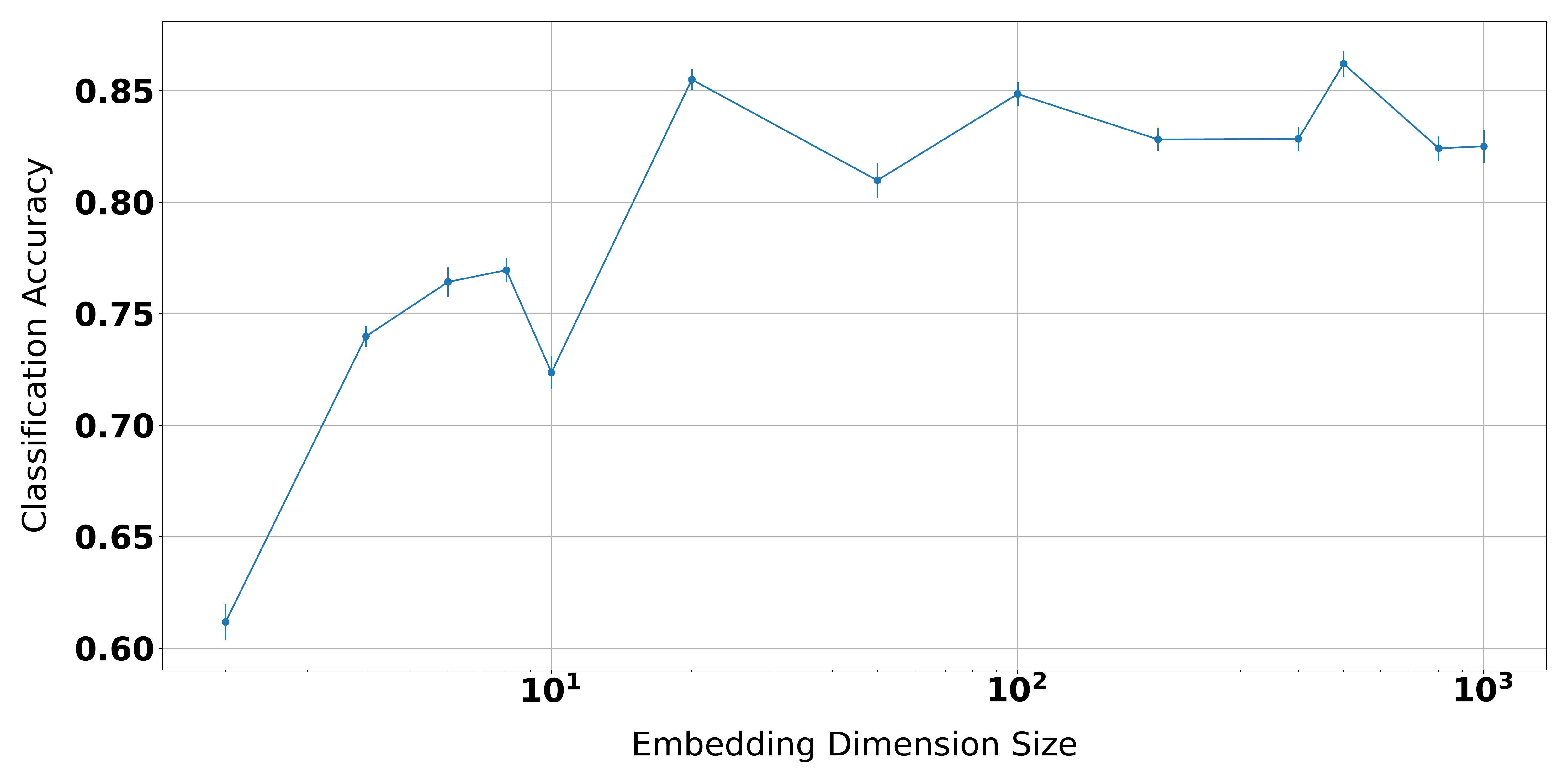}
    \caption{Classification accuracy of the SVM classifier when increasing the dimensionality of the latent projection and the number of features is fixed to 6.}
    \label{fig:high_dimensional_separability}
\end{figure}

Although the designed visualisation system is constrained to two dimensions, it is useful to consider the model's classification performance (or higher dimensional geometric separability) for sake of explainability and generalisability of the system. For this reason experiments were run to measure the SVM classification accuracy when the dimensionality of latent projection is varied between 2 and 1000 while the number of features is fixed to 6. Six features were chosen as it reflects the point where the classification accuracy when using a two-dimensional embedding deteriorates. For this reason it is of interest to see the behaviour of classifier as the dimensionality of the latent projection increases. The result of this may be seen in Figure~\ref{fig:high_dimensional_separability}. 

It can be seen that as the dimensionality of the embedding vector increases so does the model's ability to classify compounded features, however the high dimensional latent space does not enable easy visualisation of data. This being said, using higher dimensional embeddings may enable anomaly detection and aid the generative abilities of the VAE in future work. Additionally, this result further confirms the limitation of the model ability to learn more complex compounded features. 

 \begin{figure*}
        \centering
         \begin{subfigure}[c]{0.47\textwidth}
             \centering
             \includegraphics[width=0.95\textwidth]{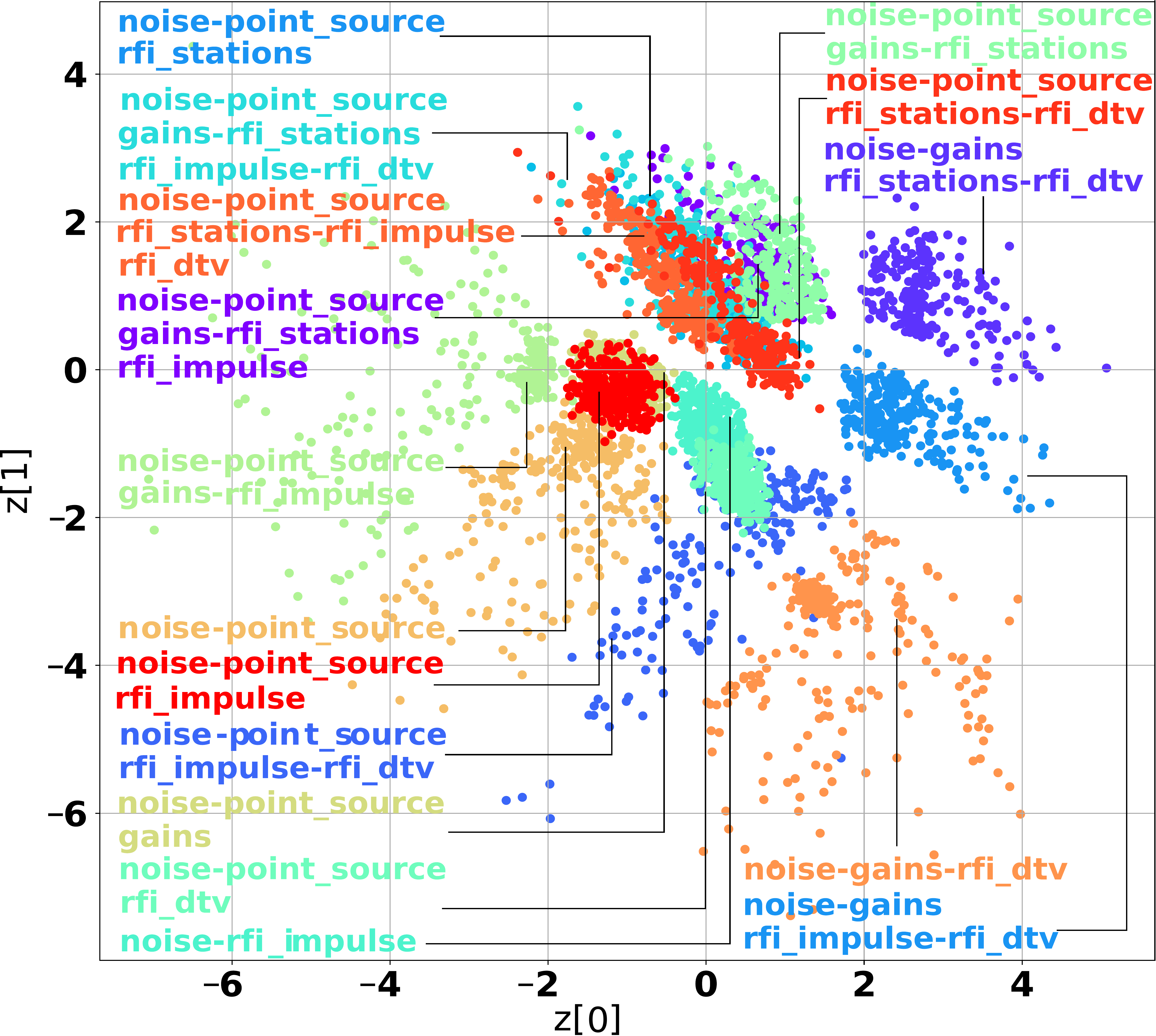}
            \caption{2D embedding of the HERA-trained VAE with the colours corresponding to the compounded labels for each feature.}
            \label{fig:Embedding_colours}
         \end{subfigure}
         \hfill
         \begin{subfigure}[c]{0.47\textwidth}
             \centering
             \includegraphics[trim={0 0 0 1.2cm},clip,width=\textwidth]{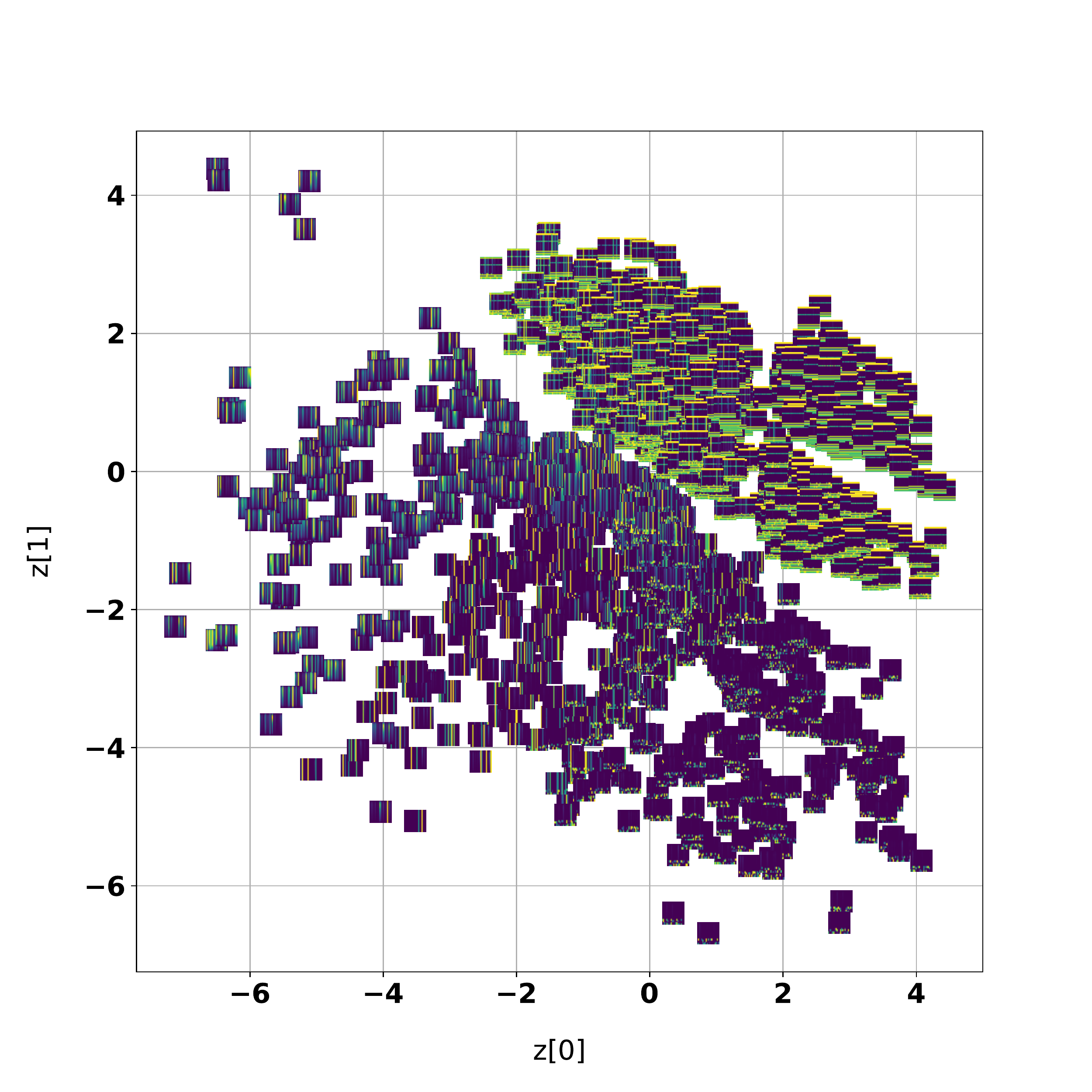}
             \caption{The embedding of the HERA-trained VAE with the magnitude of the input spectrograms superimposed onto each of the corresponding points of the embedding.}
            \label{fig:Embedding_overlay}
         \end{subfigure}
        
        \caption{The hierarchical feature separation for the magnitude and phase-based VAE trained on simulated HERA data with 6 features.}
        \label{fig:Embeddings}
\end{figure*}
    
\begin{figure}
     \centering
     \includegraphics[width=0.95\columnwidth]{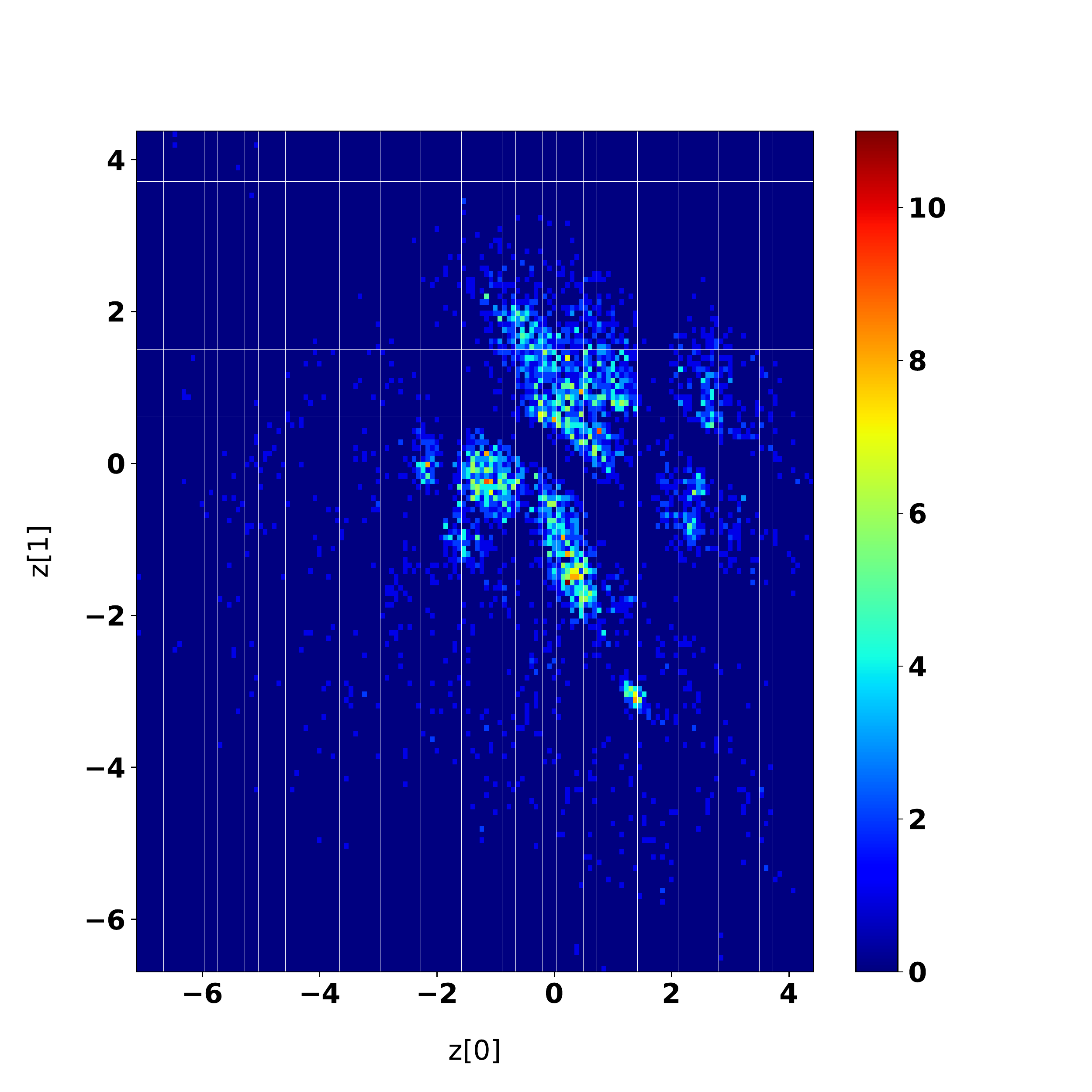}
     \caption{2D histogram showing the distribution of data within in 2D latent embedding.}
    \label{fig:Embedding_Histogram}
\end{figure}
Through the use of the magnitude and phase-based VAE that presented a feature classification accuracy of 65\% the two dimensional embedding plots shown in Figure~\ref{fig:Embeddings} were generated. It must be noted that the embeddings that yield a high classification accuracy are not presented in this work as they do not demonstrate the ideas of hierarchical feature compounding that are relevant to explanations of the embedding of the LOFAR data. For this reason and brevity sake the embeddings of the simulated HERA data with five features and less are omitted from the results. 

Figure~\ref{fig:Embedding_colours} shows the 2 dimensional embedding of of the input spectrograms with six features, namely narrow-band radio station-based and digital television broadcasting signals (RFI), broad-band impulsive RFI, gain fluctuations, source structure and Gaussian noise. In this experiment the labels associated with each compounded feature are coloured differently. 
In Figure~\ref{fig:Embedding_overlay} the same embedding is shown, except rather than colouring each point by its respective label, the magnitude component of the input spectrogram is overlaid onto each of the embedding points associated with that input. It can be seen that the features of associated with each input are clearly visually separated, additionally the segmentation of the label-coloured inputs shown in Figure~\ref{fig:Embedding_colours} is clearly reflected in this plot.

Notably, there is a clear separation in the features that contain narrow-band radio station-based RFI, shown in the top-right half, and those that do not, shown in the bottom half. Furthermore, within the station based-RFI contaminated region, there is a clear separation between the spectrograms that contain source structure, in the top left-most region, and those do not, shown in the top right-most region. Similar segmentations can be seen in the bottom-most region that do not contain radio station-based RFI, in that there is a clear separation between those spectrogram that contain sources and those that do not. 

Furthermore, the 2D histogram shown in Figure~\ref{fig:Embedding_Histogram} reflects how the input spectrograms are evenly distributed within the two dimensional embedding. It can be seen that there is a uniform distribution of data within the 2 dimensional space, with no feature occupying a single point on the grid. 

\subsection{LOFAR Data}
Here we describe the qualitative results of the model trained and evaluated on LOFAR data. This subsection uses descriptive methods to show the generalisability of the model to LOFAR data where the number of features and their compounding become more apparent. 

As shown in Section~\ref{subsec:accuracy} the inclusion of both magnitude and phase information yields improved performance. This being said, if we were to train the LOFAR-data based model on magnitude components only, then it would be expected that the embedding would appear significantly different. It is expected that the geometric separability of features such as auto-correlation will be worse, as  auto-correlations in phase appear as zeros in their corresponding spectrograms, which is an easily represented feature in the phase based embedding. 

\subsubsection{LOFAR data embedding}
\label{subsec:lofar_results}
\begin{figure*}
    \centering
         \begin{subfigure}[t]{0.49\textwidth}
         \centering
         \includegraphics[width=\textwidth]{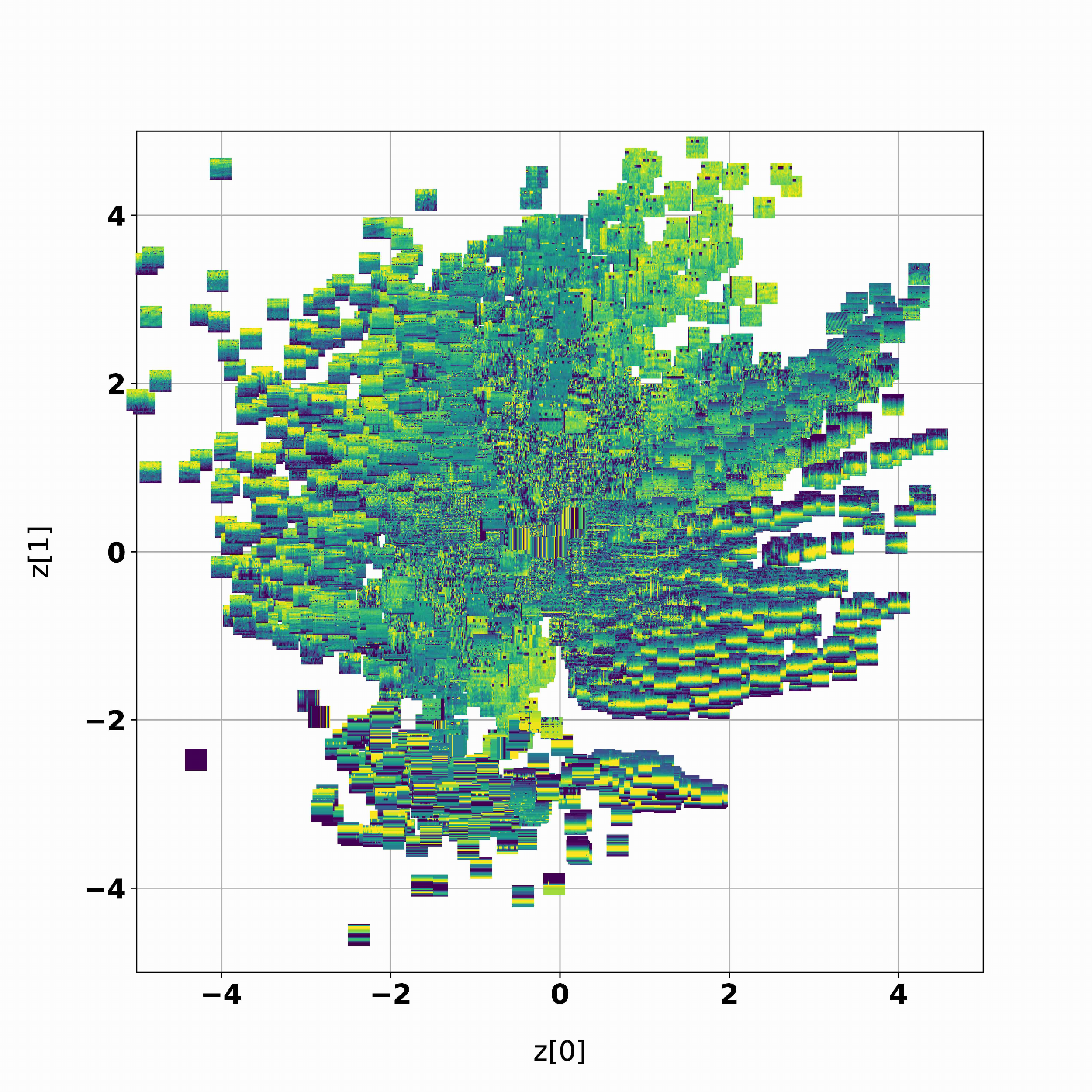}
        \caption{The embedding of the LOFAR-trained VAE when randomly sampled training data is inputted and the magnitude of each of the input spectrograms are superimposed onto each x-y coordinates generated by the encoder.}
        \label{fig:Mag_scatter}
     \end{subfigure}
     ~
     \begin{subfigure}[t]{0.49\textwidth}
         \centering
         \includegraphics[width=\textwidth]{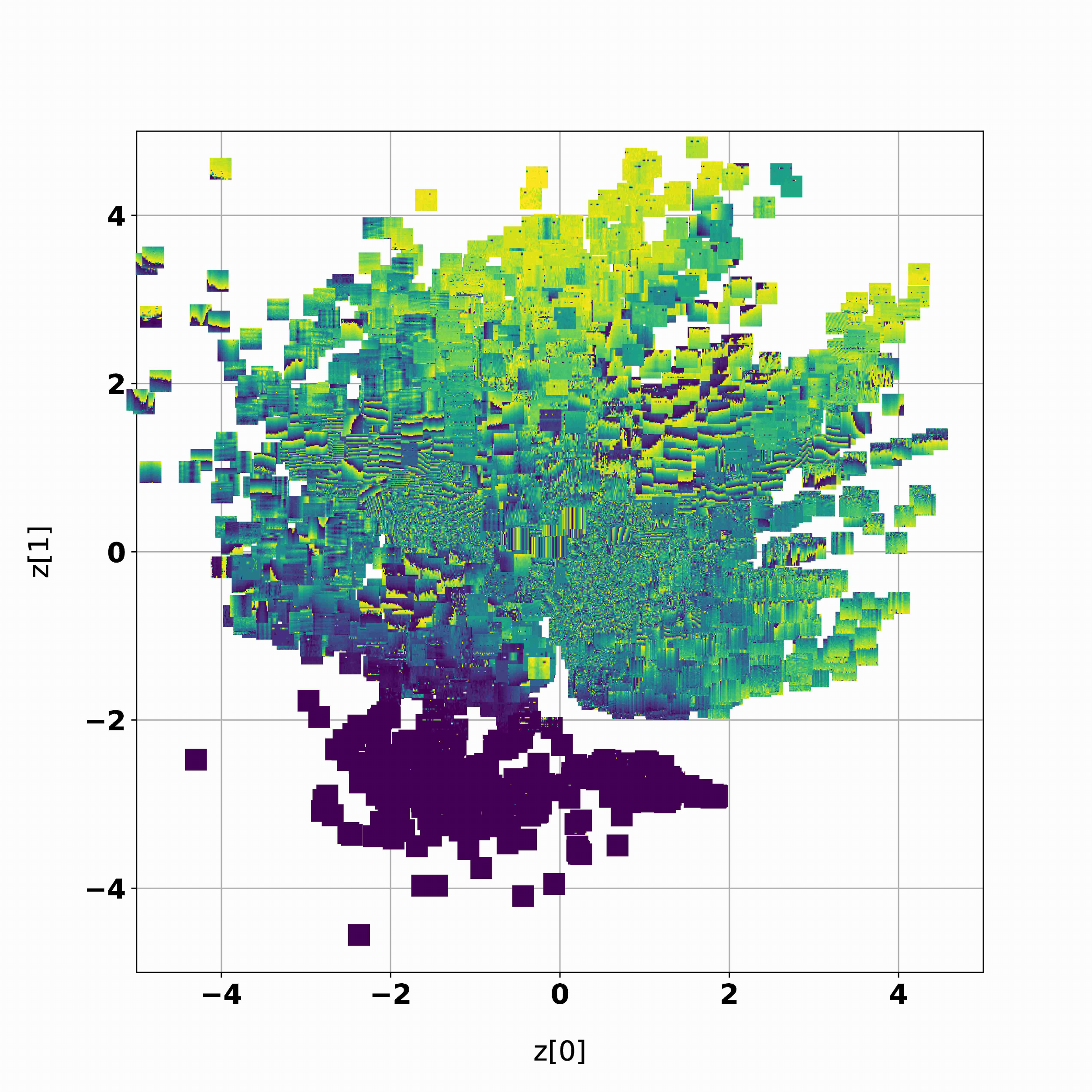}
        \caption{The embedding of the LOFAR-trained VAE when randomly sampled training data is inputted and the phase of each of the input spectrograms are superimposed onto each x-y coordinates generated by the encoder.}
        \label{fig:Phase_scatter}
     \end{subfigure}
    \caption{The embeddings of the VAE-based model for preprocessed testing data obtained from LOFAR sampled from 100 random observations. Where the magnitude and phase components are overlaid onto each point in the 2D latent space. It is shown that the model can effectively separate various source structures with an even distribution of data.}
    \label{fig:LOFAR_results}
\end{figure*}

The evaluation of the LOFAR-trained VAE was performed using 100 randomly selected  \texttt{.hdf5} files, that were not in the training set, with 256 baselines sampled from each file. Each spectrogram used for testing contained a number of features compounded in different ways. The embedding of the test data is shown in Figure~\ref{fig:LOFAR_results}, when the magnitude and phase components of the input spectrograms are superimposed on each of the points that they are projected to in the 2D latent space.

It can be seen that the generated embedding is clearly capable of grouping together visually similar features as well as geometrically separating distant features. Admittedly, the interplay between the magnitude and phase components do make interpretability of results challenging, however it is clear that the spectrograms from the test data lie on a low dimensional projection, or a manifold, with multiple vertices. 

In Figure~\ref{fig:LOFAR_results}, it can be seen that spectrograms of the auto-correlations are embedded to the bottom half of the latent space. This is shown by the normalised phase component of 0 in Figure~\ref{fig:Phase_scatter}, whereas spectrograms with high normalised phase components and \textit{block-like} RFI are placed toward the top of the embedding. Similarly, when considering the bottom-right region of Figure~\ref{fig:Mag_scatter} it can be seen that particular source structures with a high magnitude component in high frequency bands are grouped together. It can be observed that spectrograms that contain zero magnitude and phase information are geometrically separated from the manifold. This is shown in both Figures~\ref{fig:Phase_scatter} and \ref{fig:Mag_scatter} at approximately point $(-4.2,-2.5)$. 

\begin{figure*}
    \centering
     \begin{subfigure}[t]{0.45\textwidth}
         \centering
         \includegraphics[width=\textwidth]{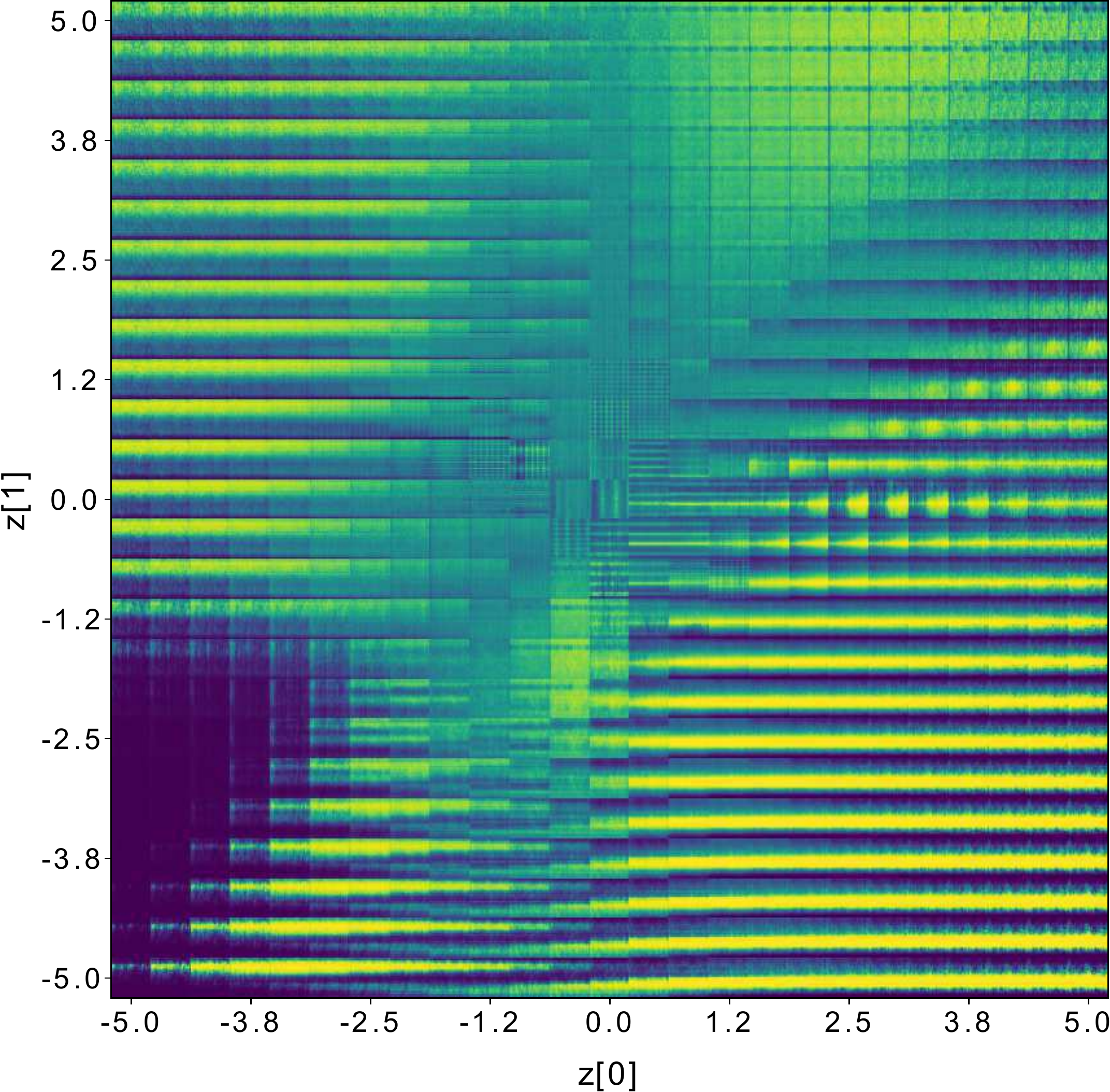}
        \caption{Magnitude embedding plot, where each point in the embedding corresponds to the learnt magnitude representations of the VAE.}
        \label{fig:Mag_embeding}
     \end{subfigure}
     ~~~~
     \begin{subfigure}[t]{0.45\textwidth}
         \centering
         \includegraphics[width=\textwidth]{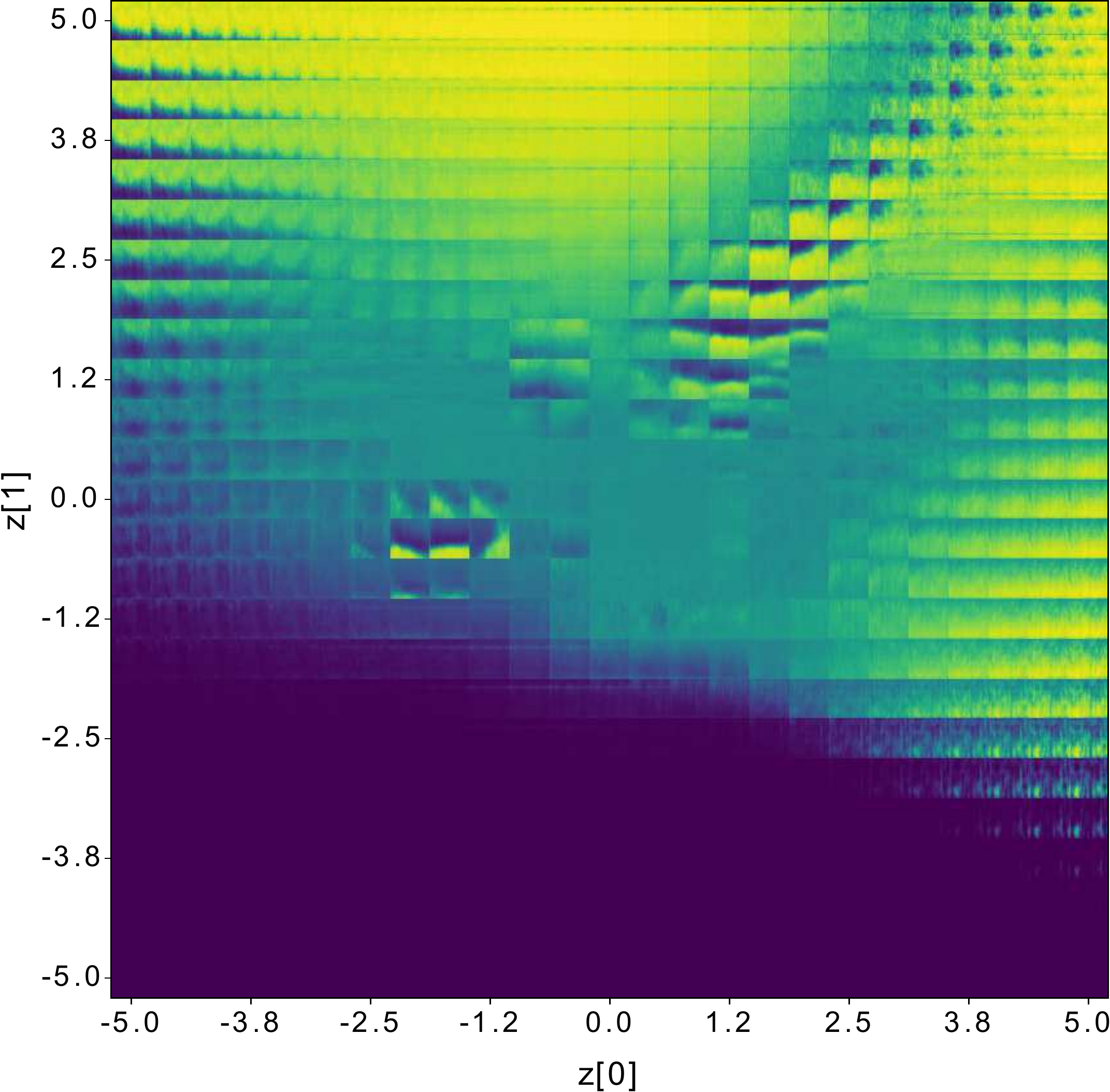}
         \caption{Phase embedding plot, where each point in the embedding corresponds to the learnt phase representations of the VAE}
        \label{fig:Phase_embeding}
     \end{subfigure}
      \hfill
     \begin{subfigure}[t]{0.45\textwidth}
         \centering
         \scalebox{1}[1.0]{\includegraphics[width=0.95\textwidth]{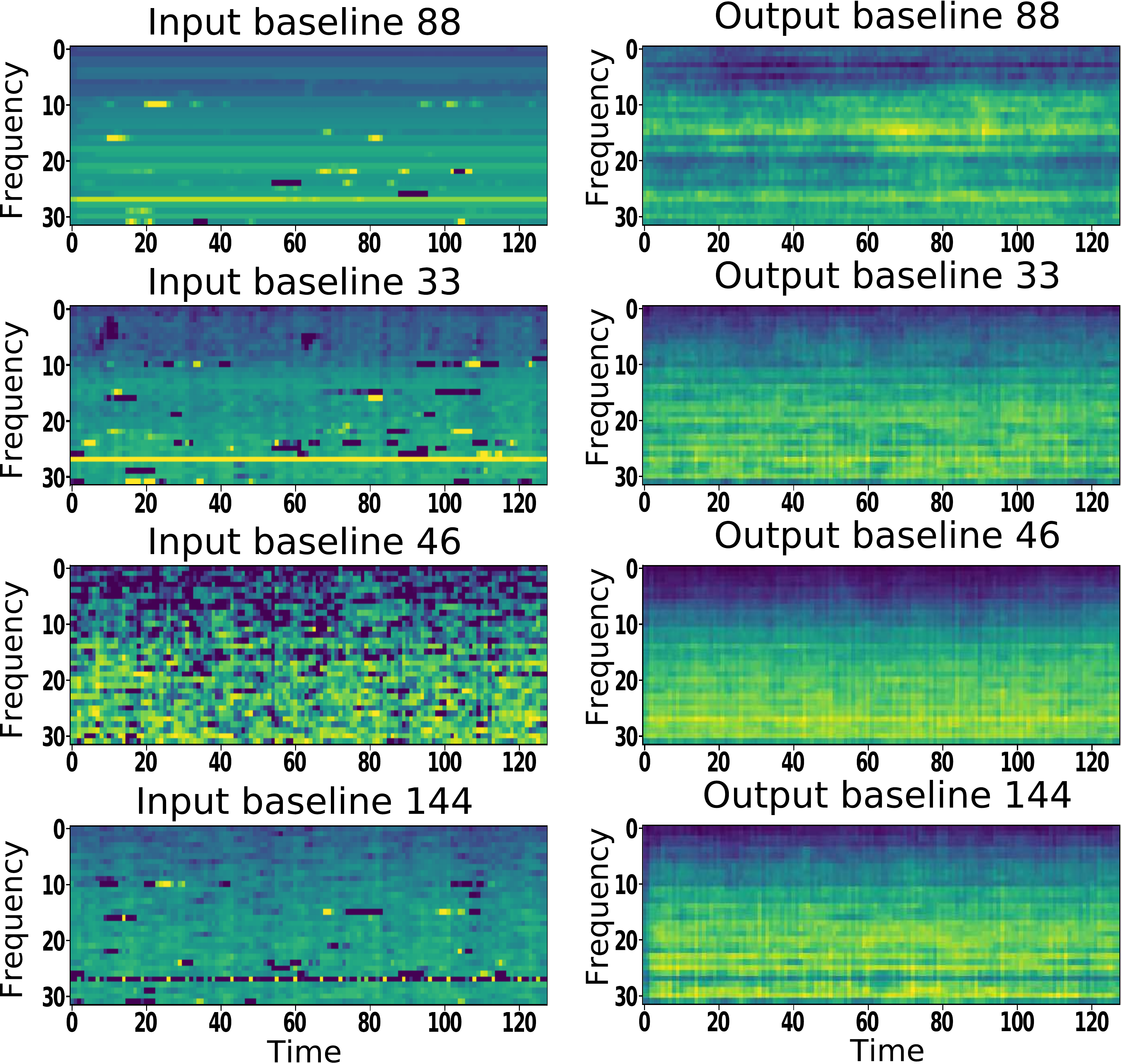}}
        \caption{The learnt magnitude representations, where the left column corresponds to the input spectrogram and the right column corresponds to the representation learnt by the VAE.}
        \label{fig:Mag_IO}
     \end{subfigure}
     ~~~
     \begin{subfigure}[t]{0.45\textwidth}
         \centering
         \scalebox{1}[1.0]{ \includegraphics[width=0.95\textwidth]{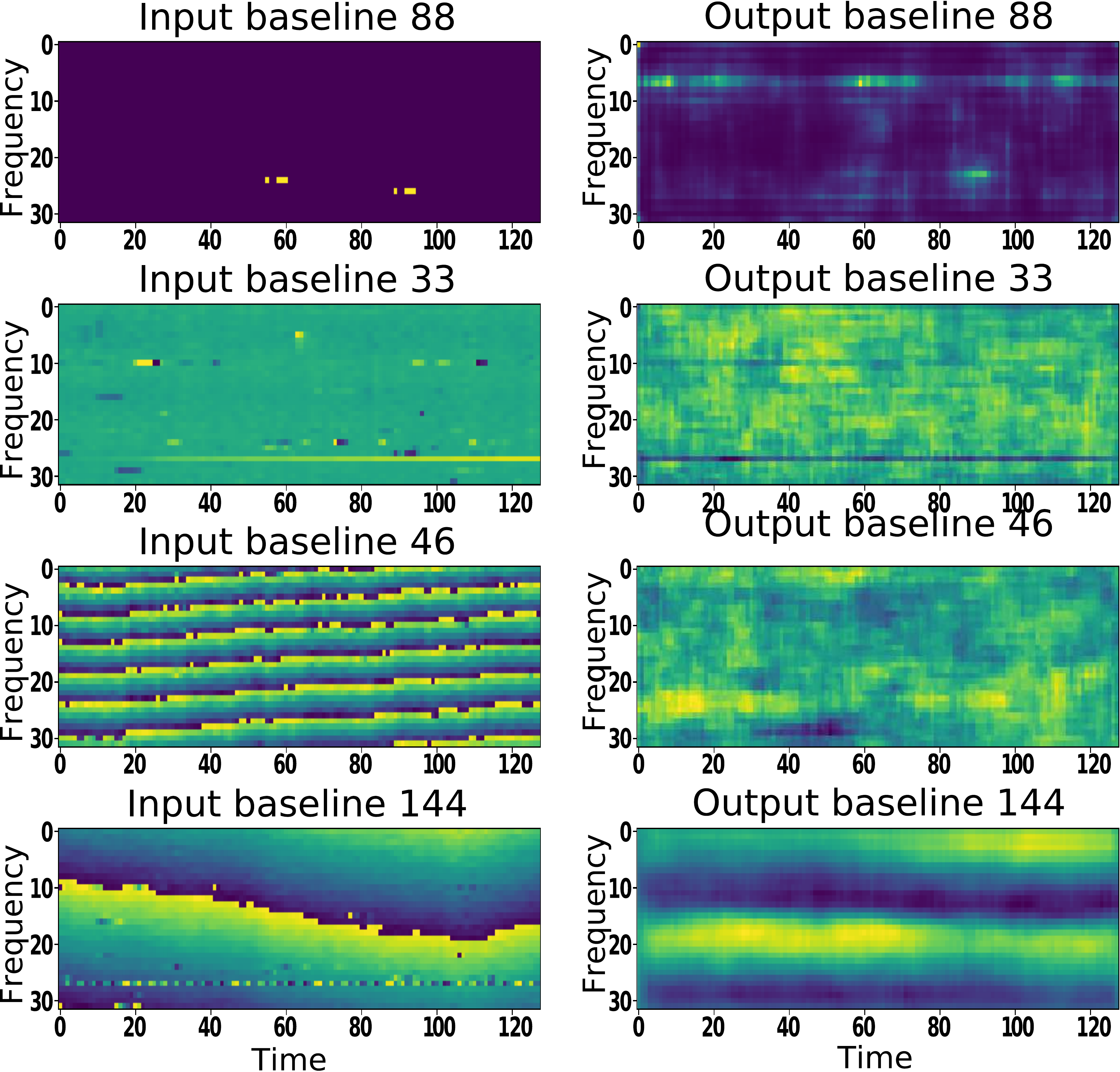}}
         \caption{The learnt phase representations, where the left column corresponds to the input spectrogram and the right column corresponds to the representation learnt by the VAE.}
        \label{fig:Phase_IO}
     \end{subfigure}
     
    \caption{The learnt magnitude and phase representations of the VAE that are superimposed onto each point of the latent space and the difference in magnitude and phase components of the input spectrograms relative to the representations that the VAE has learnt.}
    \label{fig:VAE_Embedings}
\end{figure*}

More generally, Figure~\ref{fig:VAE_Embedings} shows the decoded output of each point in the latent space superimposed onto its corresponding 2D coordinate when each spectrogram is separated into magnitude and phase components. This plot shows the geometric separability of each feature in its latent projection, which clearly reflects the results shown in Figure~\ref{fig:LOFAR_results}. 

This being said, the plots in Figure~\ref{fig:VAE_Embedings} reflect the learnt representations of the input spectrograms rather than \textit{ground-truth} of these inputs. This aspect of the generative abilities of the model is discussed in Section~\ref{subsec:generative_results}.  Figure~\ref{fig:Mag_embeding} shows clear geometric separation between various features in the latent space. It can be seen that sparse RFI-based features are grouped toward the origin of the embedding, whereas toward the top-region of the plot higher magnitude representations are learnt. 

The corresponding learnt phase-representations of the complex spectrograms reflected in Figure~\ref{fig:Phase_embeding}. It can be observed that the bottom region of the embedding corresponds to auto-correlations which have a constant 0-phase.  Furthermore, it is shown that the VAE learns that spectrograms projected around the point $(-1.7,-0.4)$ contain particular low-frequency source structure.

This result gives insight into the magnitude and phase based performance gains that were previously discussed. In this case, the lack of detail in the learnt representations are attributable to the VAE under-performing in its ability to learn phase representations of the data. However the lack of detail in the learnt phase representations are to be investigated in future work. 

% trim={<left> <lower> <right> <upper>}

\subsubsection{Generative abilities of LOFAR trained model}
\label{subsec:generative_results}
\Cref{fig:Mag_IO,fig:Phase_IO} show the differences between input spectrograms separated into their respective magnitude and phase components and the learnt magnitude and phase representations of the VAE. These figures clearly show the limitations of the network to accurately learn the correct representations of the LOFAR data. It can be observed that the network has learned more refined representations of the phase component of the spectrograms as seen in Figure~\ref{fig:Phase_IO}. 

The top-most plot of Figure~\ref{fig:Phase_IO} illustrates that the model learns a \textit{general} structure of the input. This is reflected in the correct dynamic range scaling, however finer detail features in the input spectrogram located at time-samples 60 and 80 are not regenerated. Furthermore, it can be seen in the bottom-most plot that the network seems to learn a \textit{blurred} sinusoidal representation of the input.

An easy way to improve the performance of this model would be to increase the number of dimensions in the latent vector, however this would defeat the purpose of this work. 

Interestingly, Figure~\ref{fig:Mag_IO} shows how the VAE learns the intensity of the magnitude component of the spectrograms, however fails to learn more chaotic features such as spurious RFI. In contrast, the phase representation, shown in Figure~\ref{fig:Phase_IO}, the model seems to learn the narrow band RFI in the input spectrogram from the 33rd baseline.

As already mentioned, the learnt source representations are clearly incorrect, yet they show convincing uses for the generative abilities of the VAE to correctly embed the spectrograms obtained from LOFAR. 

\subsubsection{Prototype data inspection environment}
\begin{figure*}
\centering
\includegraphics[width=1\linewidth]{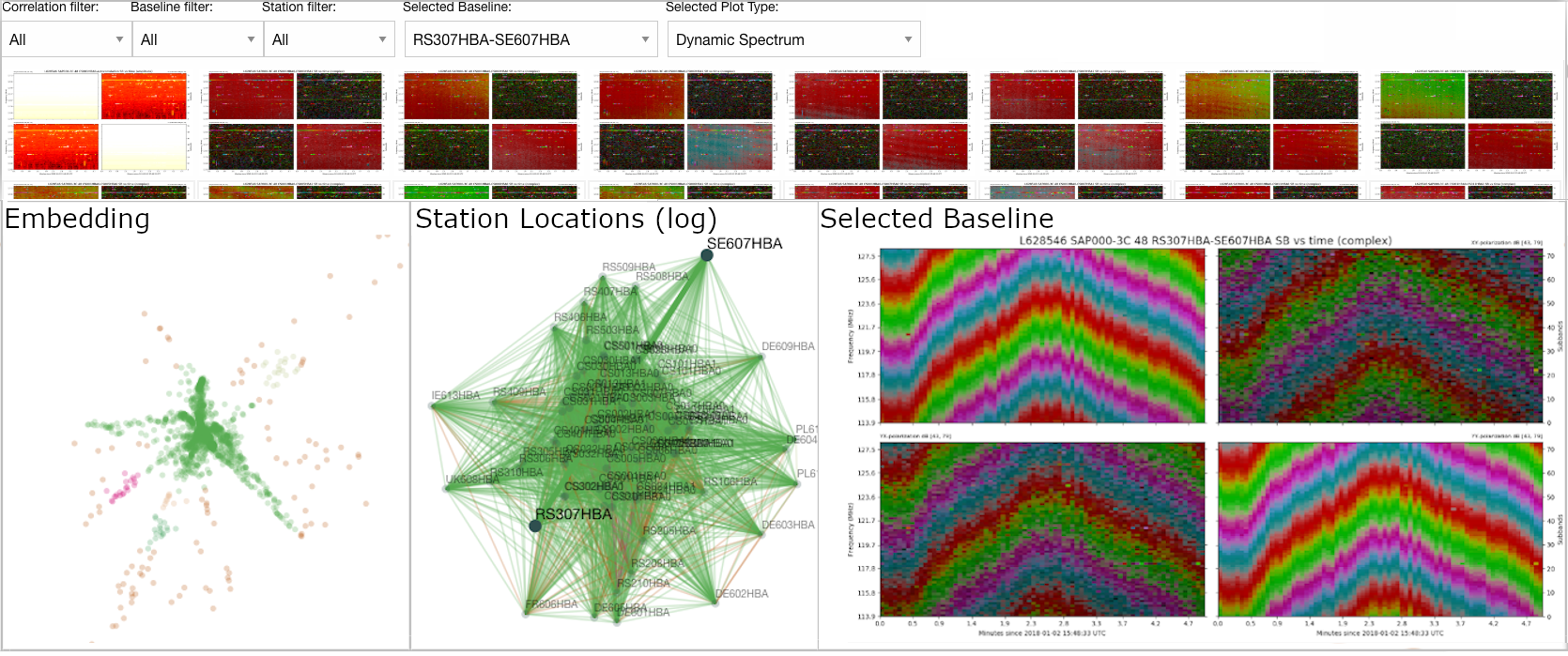}
\caption{Screenshot of the prototype-web interface being used at the ASTRON observatory. The top-most selection boxes correspond to the filters on the baselines and the embedding that will be displayed. The row of images below the filters correspond to all spectrograms within a particular filter. The left-most image in the bottom row corresponds to the embedding of the particular observation with the filters applied, whereas the centre plot on the bottom row is the log-scale station location plot. Finally the right-most plot on the bottom row is the spectrogram corresponding to the baseline selected in the embedding. }
\label{fig:adder}
\end{figure*}

We integrated the VAE into a flask-based \footnote{https://flask.palletsprojects.com/en/1.1.x/} web-application to make the model accessible to astronomers. A screen-shot of the interface is shown in Figure~\ref{fig:adder}. The web-interface displays the embedding of a selected observation. Moreover, it enables the filtering of displayed results based on various criteria such as correlation-type, station location and others. 

The interface enables operators to use the embedding as a diagnostic tool. They can detect outliers in the embedding and trace the errors back to a particular station. The system is currently being evaluated at the ASTRON observatory.

\section{Conclusions and Future Work}
\label{sec:Conclusions}
    Deep learning has proven to be an invaluable tool for many different fields and its usefulness in radio astronomy is becoming readily apparent. In this work we have shown that machine learning techniques have influenced many processes within the radio astronomy pipeline, and we predict that this trend is only going to increase due to the increasing data rates and complexity from future radio telescopes.

    We showed that a deep convolutional VAE that uses both magnitude and phase components of the complex visibilities is capable of producing low dimensional embeddings to aid astronomers with data quality analysis and system health management. 
    
    We proved that, through the use of a simple preprocessing pipeline and a relatively small amount of data (compared to the amounts being generated on a daily basis by LOFAR), our VAE is capable of learning low dimensional embeddings of the high dimensional features from both simulated HERA data as well as real LOFAR data. 
    
    We quantitatively proved how the various parameters of these models effect the performance in geometric separability of data obtained from a constrained simulation from of the HERA telescope. We showed the limits of the model by considering the dimensionality of the projections and the accuracy of a na\"ive SVM classification algorithm. Through this experimentation, we showed that this model is capable of generalising to the \emph{unlabelled} real data from LOFAR.
    
    The VAE and SVM combination scores between 65\% and 90\% accuracy, and the geometric separability of features in two dimensions follows a hierarchical compounding scheme. This combination is proven a useful method for assisting human operators to diagnose failure. We showed the integration of the model into a diagnostic web-framework enables telescope operators to pinpoint system failures. 
    
    We do recognise that the model is limited by its generative abilities to learn the complex compounded features obtained from the LOFAR telescope. We show that there is a trade-off when trying to project the data to a two dimensional latent space and maintain the high fidelity generative abilities of the VAE. We believe that this limitation can be resolved through the use of more sophisticated data augmentation and preprocessing techniques which decouples the compounding of features present in the time-frequency domain. Some examples of techniques that could be used are wavelet scattering transforms~\cite{Bruna2013}, 2D Fourier analysis~\cite{Yin2019} and autoencoder based data augmentation~\cite{Ralph2019}.
    
    Additionally, a more sophisticated GAN based models could be used to overcome the learning short-comings of the VAE as they have been shown to be more effective in their abilities to generate new samples from a learnt distribution of data~\cite{Akcay}. Furthermore, we would like to investigate the possibility of training an ensemble of networks, each receiving different bands of the astronomical data such as the LBA and HBA bands of the LOFAR telescope. This might aid both the generative abilities and the embedding quality of a single network, as there would be less variation of features in a single training set.
    
    In future work the anomaly detection abilities of a VAE based system will be investigated through use of higher dimensional latent projections. We plan to integrate this visualisation tool into the LOFAR observatory pipeline, which may grant the opportunity to investigate active learning in the context of LOFAR. 
 
    The github repository containing the respective code of the system can be found at \hyperlink{https://github.com/mesarcik/DL4DI}{https://github.com/mesarcik/DL4DI}~\cite{Mesarcik2020}.

\section*{Acknowledgements}

 This work is part of the "Perspectief" research programme "Efficient Deep Learning" (EDL, \href{https://efficientdeeplearning.nl}{https://efficientdeeplearning.nl}), which is financed by the Dutch Research Council (NWO) domain Applied and Engineering Sciences (TTW), and the Innovatie Cluster Drachten (ICD) project ML\&AI \href{https://www.icdrachten.nl}{https://www.icdrachten.nl}. The research makes use of  radio astronomy data from the LOFAR telescope, which is operated by ASTRON (Netherlands Institute for Radio Astronomy), an institute belonging to the Netherlands Foundation for Scientific Research (NWO-I). We acknowledge the use of Python software especially the \href{https://www.tensorflow.org/}{tensor-flow}, \href{https://keras.io/}{keras} and \href{https://numpy.org/}{numpy} in our data analysis. This research also made use of the  \href{https://github.com/HERA-Team/hera_sim}{HERA radio astronomy simulator}.
We also acknowledge Ronald Nijboer (ASTRON, currently Netherlands Defence Academy), Giovanni Mariani (Qualcomm), and Hanno Spreeuw (Netherlands eScience Center) for their help in setting up the research directions at the start of the project. And thanks to Michiel Brentjens, Jorrit Schaap and Thomas Franzen (ASTRON) for their support in defining and implementing the data compression and providing LOFAR data.

\bibliographystyle{mnras}
\bibliography{main.bib} 

% \input{appendix/Appendix.tex}

%%%%%%%%%%%%%%%%%%%%%%%%%%%%%%%%%%%%%%%%%%%%%%%%%%
%%%%%%%%%%%%%%%%%%%%%%%%%%%%%%%%%%%%%%%%%%%%%%%%%%
% Don't change these lines
\bsp	% typesetting comment
\label{lastpage}
\end{document}